12,117 words in the main text

196 words in the abstract

65 references

14 tables and 8 figures in the main text

7 table in the Appendix

# Occlusion-induced risk and interventions in pedestrian-autonomous truck interactions on multi-lane roads: A virtual reality study

Yun Ye[1,2], Yuan Che[2], S.C. Wong[3], Stergios-Aristoteles Mitoulis[1], Haoyang Liang[4,5†]

[1] Centre for Global Infrastructure Resilience, The Bartlett School of Sustainable Construction, University College London, London, United Kingdom

[2] Faculty of Maritime and Transportation, Ningbo University, Ningbo, China

[3] Department of Civil Engineering, The University of Hong Kong, Hong Kong, China

[4] College of Transport and Communications, Shanghai Maritime University, Shanghai, China

[5] College of Transportation, Tongji University, Shanghai, China

† **Correspondence to**: hyliang@shmtu.edu.cn

**Acknowledgement:** This research was supported by the National Natural Science Foundation of China (Project No. 72501150, 52302378), the Zhejiang Provincial Natural Science Foundation of China (Grant No. LQN25E080011), the Ningbo Natural Science Foundation (Grant No. 2024J440), the National "111" Centre on Safety and Intelligent Operation of Sea Bridges (Project No. D21013), and the Research Grants Council of the Hong Kong Special Administrative Region, China (Project No. T32-707/22-N). The third author was also supported by Francis S Y Bong Professorship in Engineering. The funders had no role in the study design, data collection or analysis, manuscript preparation, or the decision to publish.

**ABSTRACT**

Autonomous trucks (ATs) may introduce distinct pedestrian-safety risks because of their large physical dimensions, constrained braking capability, limited driver-based communication cues, and potential to occlude surrounding traffic. This study employed a controlled virtual reality experiment with 54 participants to investigate pedestrian–AT interaction risk in an unsignalized multi-lane crossing scenario and to evaluate occlusion-targeted risk mitigation strategies. The experiment examined the effects of near-side vehicle type, weather condition, and far-side vehicle yielding strategy on pedestrian behavior, perceived risk, and objective safety. Based on a representative high-risk scenario, three targeted interventions were designed and tested: an environment-aware external human–machine interface (eHMI), a projected eHMI, and an auditory warning. The results showed that ATs increased perceived risk and encouraged more cautious crossing behavior, suggesting a risk-compensation effect. However, this compensation was weakened under rainy conditions, where braking-related safety margins were reduced. AT-induced occlusion further increased far-side interaction risk by limiting pedestrians' recognition of hidden vehicles. Among the three interventions, the projected eHMI showed the best overall performance, improving objective safety margins, enhancing risk awareness, and supporting behavioral adjustment. These findings highlight the need for AT-specific interface and warning strategies that address both intention communication and risk localization.

*Keywords*: Autonomous trucks; external human–machine interface; pedestrian safety; virtual reality; occlusion-induced risk; targeted intervention.

## 1. Introduction

As automated driving technology expands from passenger vehicles to freight and logistics operations, autonomous trucks (ATs) are moving toward open-road deployment (Tsugawa et al., 2016; Noruzoliaee et al., 2021). This transition raises new pedestrian-safety concerns in freight-related environments, including trunk roads, logistics corridors, and open-road segments connecting freight facilities. These scenarios are particularly relevant in China, where road freight remains a dominant component of the transport system and autonomous commercial vehicles are expected to develop rapidly in logistics applications (Ministry of Transport of the People's Republic of China, 2024; LeadLeo Research Institute, 2024). Pedestrians in such environments may encounter uncontrolled crossing situations near roadside communities, industrial areas, or logistics facilities, where crossing decisions depend strongly on interactions with surrounding vehicles. Similar uncertain interaction contexts have been examined in pedestrian-AV studies, and pedestrians in China have also been observed making sequential crossing decisions on unmarked roads under complex traffic conditions (Palmeiro et al., 2018; de Clercq et al., 2019; Zhuang and Wu, 2011; Feng et al., 2024). Therefore, understanding pedestrian–AT interaction safety in open-road crossing environments is of practical importance.

Compared with passenger autonomous vehicles (AVs), ATs introduce distinct interaction risks because they combine truck-specific physical characteristics with automation-related uncertainty (Ye et al., 2026a). Their larger dimensions, higher potential crash severity, constrained braking capability, and reduced availability of driver-based communication cues may affect pedestrian risk perception, crossing decisions, and objective safety margins (Roudsari et al., 2004; Schubert et al., 2023; Velasco et al., 2019). More importantly, AT-related risk may operate through two interconnected pathways. The first is direct pedestrian–truck interaction risk, in which pedestrians respond to the truck's size, motion, braking capability, and communication ambiguity. The second is occlusion-induced risk, in which the truck body restricts pedestrians' visibility of adjacent traffic and may conceal far-side vehicles or conflict areas (Chen et al., 2025). Although occlusion-related risk has long existed in conventional truck operations (Fabricius et al., 2022), automated and connected vehicle technologies create new opportunities to mitigate such risk by detecting hidden hazards and communicating relevant information to pedestrians.

Existing external Human–Machine Interfaces (eHMIs) have mainly been investigated in pedestrian–AV interaction because they can support pedestrians' interpretation of vehicle awareness, motion state, yielding intention, and crossing opportunities (De Clercq et al., 2019; Tabone et al., 2023; Ye et al., 2026b). However, these interfaces primarily communicate ego-vehicle information and may be insufficient in AT-related occlusion scenarios, where the critical hazard may originate from hidden vehicles or conflict areas beyond the truck itself. Accordingly, AT-oriented interventions need to extend conventional intention communication toward occlusion awareness, hidden-risk recognition, and spatial risk localization. Auditory warnings may also provide a complementary non-visual channel by attracting pedestrian attention to potential hazards (Soares et al., 2021; Bindschädel et al., 2023).

These issues are particularly important in multi-lane crossings, where AT-related risk may transfer from direct interaction with the near-side vehicle to subsequent interaction with far-side vehicles. Environmental conditions such as rain may further reduce braking performance and compress objective safety margins, while the far-side vehicle's yielding strategy may influence pedestrians' risk recognition and crossing decisions after passing the occluding vehicle. Therefore, it is necessary to clarify how AT interaction risk is formed in multi-lane crossing scenarios and whether occlusion-targeted interventions can improve risk recognition, behavioral adjustment, and objective safety.

To address these issues, this study conducted a controlled virtual reality (VR) experiment to examine pedestrian–AT interaction risk and occlusion-aware risk mitigation in a multi-lane crossing scenario (Ye et al., 2020; Ye et al., 2023; Wong et al., 2023). The main contribution of this study lies in proposing and empirically validating an occlusion-aware risk mitigation strategy for pedestrian–autonomous truck interaction. Moving beyond conventional pedestrian–AV communication approaches that primarily convey vehicle awareness or yielding intention, this study focuses on supporting occlusion awareness, hidden-risk recognition, and spatial risk localization under truck-induced occlusion conditions. Specifically, the experiment first analyzed how near-side vehicle type, weather condition, and far-side AV interaction strategy affected pedestrian behavior, perceived risk, and objective safety, thereby identifying both direct pedestrian–AT risk and occlusion-induced secondary risk. Based on the representative high-risk scenario, three targeted interventions (i.e., environment-aware eHMI, projected eHMI, and auditory warning) were designed and tested. Their effectiveness was evaluated using perceived risk, pedestrian behavior, deceleration to safety time (DST), and improved time advantage (ITADV) to assess whether, and to what extent, occlusion-related information could support behavioral adjustment and improve objective safety margins.

The remainder of this paper is organized as follows. Section 2 reviews related studies on AT risk attributes, pedestrian interaction risk, and safety interventions. Section 3 describes the VR experiment and statistical methods. Section 4 presents the results of risk formation and intervention-effect experiments. Section 5 discusses the findings and practical implications, and Section 6 concludes the paper with limitations and future research directions.

## 2. Literature Review

### 2.1 Risk formation in pedestrian-ATs interaction

Compared with passenger AVs, ATs create a distinct risk context because pedestrian safety is influenced by both truck-specific physical characteristics and automation-related attributes. These risks can be broadly classified into two interconnected pathways: direct pedestrian–truck interaction risk and occlusion-induced secondary risk involving surrounding traffic.

Direct interaction risk arises from the physical and communicative characteristics of ATs. Large vehicle dimensions increase collision severity and may impose stronger visual pressure on pedestrians before crossing (Fabricius et al., 2022; Roudsari et al., 2004; Schubert et al., 2023). In addition, ATs generally have greater mass and inertia than passenger vehicles, resulting in more constrained deceleration and stopping capability, particularly under low-adhesion conditions such as rain (Sun et al., 2015; Jung et al., 2014; Jiang et al., 2024). Unlike human-driven vehicles, ATs also lack implicit communication cues such as eye contact, gestures, and body posture, which may increase uncertainty regarding vehicle awareness and yielding intention (Velasco et al., 2019; Palmeiro et al., 2018). These factors jointly influence pedestrians' risk perception and crossing decisions (Ye et al., 2026a; Jayaraman et al., 2019; Zhou et al., 2023; Feng et al., 2024).

Beyond direct interaction, ATs may generate additional risk by restricting pedestrians' access to traffic information. Due to their large geometry and structural configuration, trucks can create substantial occlusion zones and reduce the visibility of nearby vulnerable road users and adjacent traffic (Chen et al., 2025). Previous studies have shown that truck-related conflicts frequently occur during turning, parallel movements, and visibility-restricted situations (Schindler and Piccinini, 2021; Schubert et al., 2023; Mole and Wilkie, 2017). From the pedestrian perspective, however, occlusion is not only a detection issue but also an information limitation problem, as the truck body may conceal approaching vehicles and conflict sources. Therefore, in multi-lane crossing scenarios, ATs may simultaneously function as the interaction counterpart and an occluding object that creates hidden-risk conditions.

Although previous studies have identified several AT-related risk attributes, their combined influence on pedestrian interaction safety remains insufficiently understood. In particular, how ATs affect pedestrians' direct interaction with the truck while simultaneously altering their perception of surrounding traffic through occlusion requires further investigation.

### 2.2 Behavioral and risk indicators in pedestrian-AVs interaction

Previous pedestrian-AV studies have commonly used behavioral indicators, including waiting time, crossing initiation, accepted gap size, crossing time, walking speed, and crossing success rate, to describe crossing responses (Velasco et al., 2021; Figueroa-Medina et al., 2023; Song et al., 2023; Man et al., 2025). These indicators capture decision-making and movement execution, but they do not directly quantify whether sufficient safety margin remains. Therefore, surrogate safety measures are needed to evaluate objective risk from pedestrian-vehicle spatiotemporal relationships. Time-to-collision (TTC), post-encroachment time (PET), safety margin, and related temporal conflict measures have been widely used near potential conflict areas (Petzoldt, 2014; Che et al., 2026), and PET-related indicators have also been applied to pedestrian-AV safety assessment (Chen et al., 2025).

Different indicators capture different conflict dimensions. TTC measures the remaining time before a potential collision under assumed motion conditions, whereas PET measures temporal separation between road users passing through the same conflict area (Ward et al., 2015; Chen et al., 2017). For pedestrian-AT interaction, braking-related constraints and occlusion-related time advantage also need attention. Deceleration to safety time (DST) relates required deceleration to actual deceleration capacity and is therefore suitable for braking-related objective risk in AT scenarios (Hupfer, 1997). Improved time advantage (ITADV) describes the relative arrival-time relationship between the pedestrian and the vehicle while incorporating a minimum safety time threshold, making it useful for assessing temporal redundancy under restricted visibility (Sheng et al., 2025). Behavioral indicators and surrogate safety measures provide complementary perspectives, but pedestrian-AT studies require indicators that capture both braking-related vehicle dynamics and occlusion-related time advantage.

### 2.3 Interventions for pedestrian-AVs interaction safety

Because AVs lack implicit driver-related communication cues, external communication and warning measures have been widely examined to support pedestrian-vehicle interaction. Conventional eHMIs typically display the vehicle's own status, awareness, or yielding intention through text, icons, light bands, or other vehicle-mounted interfaces (Lyu et al., 2024; de Winter and Dodou, 2022). De Clercq et al. (2019) found that eHMIs can affect pedestrians' crossing decisions, while Deb et al. (2018) showed that external messages can improve pedestrians' understanding of AVs and perceived interaction safety. However, eHMI effectiveness depends on design features and traffic context. Lee et al. (2022) noted that pedestrians' interpretation of novel eHMIs is influenced by vehicle kinematics and interface familiarity, and Ye et al. (2026b) showed that eHMIs may interfere with pedestrian behavior on multi-lane streets. This context dependence may be stronger in AT scenarios, where pedestrians need information about both the truck's own state and occluded traffic.

Projected eHMIs provide another visual channel by placing information in the road environment rather than on the vehicle surface. They can present warnings near potential conflict areas and align information with pedestrians' spatial perception. Shah et al. (2022) found that projection distance and contrast affect pedestrians' acceptance of projected warnings in visibility-restricted scenarios, while Tabone et al. (2023) indicated that spatially mapped information can improve understanding of risk areas and vehicle intent. Auditory warnings provide a non-visual channel for attracting attention to traffic risks and may be useful under restricted visibility or high attentional load. Soares et al. (2021) found that visual and auditory cues can affect pedestrians' crossing decisions, while Bindschädel et al. (2023) showed that auditory signals can supplement eHMIs and vehicle kinematics in communicating AV intent. However, auditory warnings often have limited spatial directivity and may provide less support for identifying the precise risk source or conflict area (Ulrich et al., 2014).

In passenger-AV contexts, these measures mainly help pedestrians interpret vehicle awareness, yielding intention, and crossing opportunities. AT scenarios may require broader support because large vehicle bodies, visual occlusion, and braking constraints affect observation and safety judgment. In multi-lane crossings, a near-side AT may also block pedestrians' view of adjacent-lane vehicles, creating occlusion conditions and hidden conflicts. Therefore, interventions for AT scenarios should be considered not only as intention-communication tools, but also as mechanisms for occlusion awareness, hidden-risk recognition, and risk localization. In AT-related occlusion scenarios, it remains unclear whether these measures can support occlusion awareness, hidden-risk recognition, spatial risk localization, and objective safety improvement.

## 3. Methods

This study adopted a controlled VR design consisting of two research stages. These stages refer to the logical structure of the study design and analysis rather than to two chronological experimental blocks. Stage 1 focused on risk formation and examined pedestrian crossing behavior under different combinations of near-side vehicle type, weather condition, and far-side AV interaction strategy without additional intervention measures. Stage 2 focused on intervention validation and evaluated whether targeted interventions could mitigate risk in a predefined representative high-risk scenario. This representative high-risk scenario was defined during the experimental design stage because it combined strong truck-induced occlusion, adverse weather-related braking constraints, and a non-yielding far-side vehicle. All risk-formation and intervention trials were presented as a single randomized set to reduce order and learning effects. The framework is shown in Fig. 1.

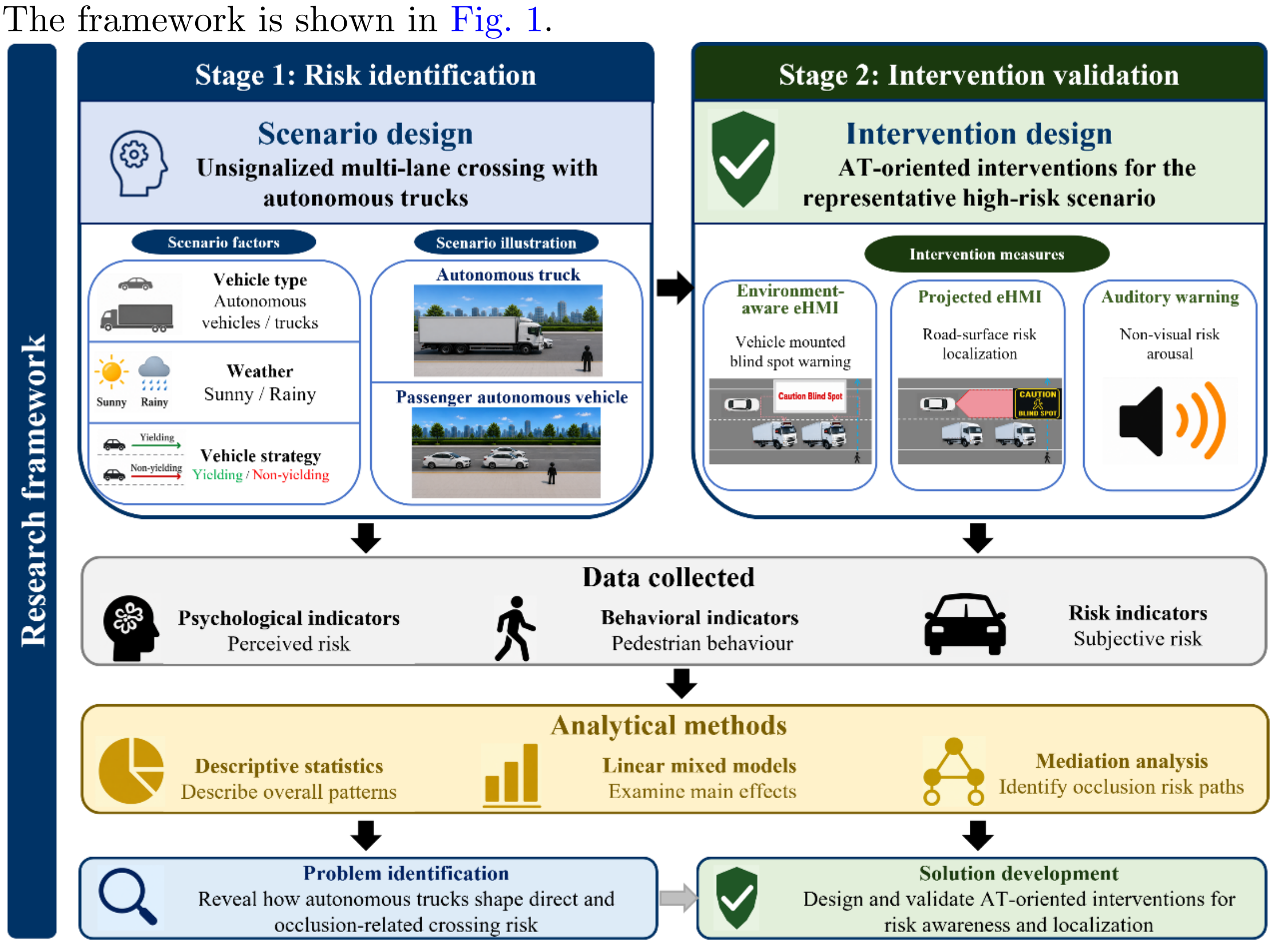

**Fig. 1.** An overview of the research framework.

### 3.1 Experiments: platform and apparatus

The experiments were conducted on a VR platform developed based on the immersive Cave Automated Virtual Environment (CAVE) Laboratory at Tongji University (https://tops.tongji.edu.cn/jxpy/sypt.htm) The platform integrated head-mounted VR devices, interactive movement control, dynamic vehicle simulation, and synchronous data recording, enabling pedestrian-vehicle interactions to be reproduced under controlled and repeatable conditions.

As shown in Fig. 2(a), participants entered the virtual road environment through a head-mounted display and used a handheld controller to control movement and viewpoint adjustment. The controller allowed participants to initiate, pause, and resume movement during the crossing task, and identical locomotion settings were maintained across all experimental conditions to minimize systematic variation associated with the movement mode. This setup allowed participants to complete the crossing task from a first-person pedestrian perspective while researchers monitored the experiment and managed task operation through a workstation. Fig. 2(b) shows the first-person view presented to the participants, including the multi-lane roadway, roadside environment, surrounding elements, and approaching AVs. The same VR scene was used to present vehicle type, weather condition, interaction strategy, and intervention information across experimental conditions.

The pedestrian simulation system used an HTC VIVE Focus 3 stereoscopic head-mounted display, with a total resolution of 4896 × 2448 pixels, 2448 × 2448 pixels per eye, and a refresh rate of 90 Hz. For auditory-warning conditions, the warning sound was delivered through the headset. The workstation used for experiment control and data management was equipped with an Intel Core i7-14700K processor and an NVIDIA RTX 4070 Ti graphics card with 12 GB of video memory. During the experiment, the platform continuously recorded pedestrian movement, vehicle trajectories, and pedestrian-vehicle spatiotemporal relationships for subsequent behavioral and objective risk analyses.

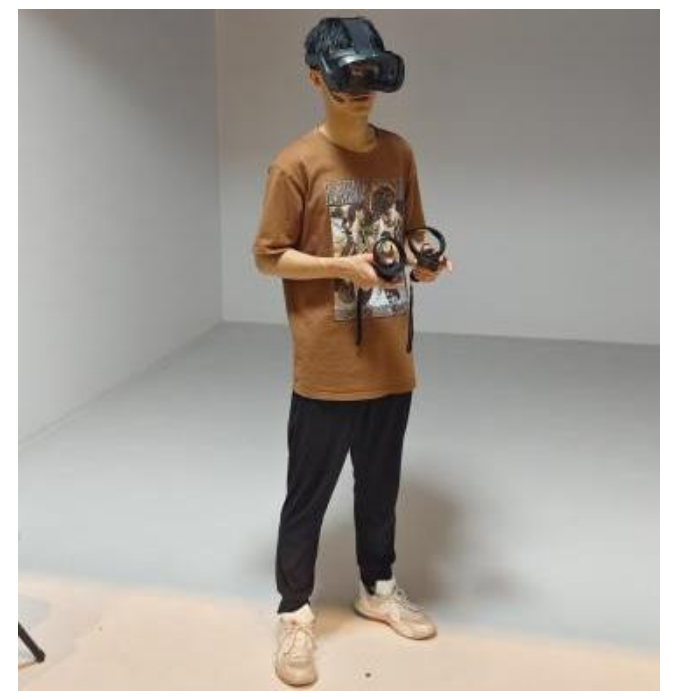
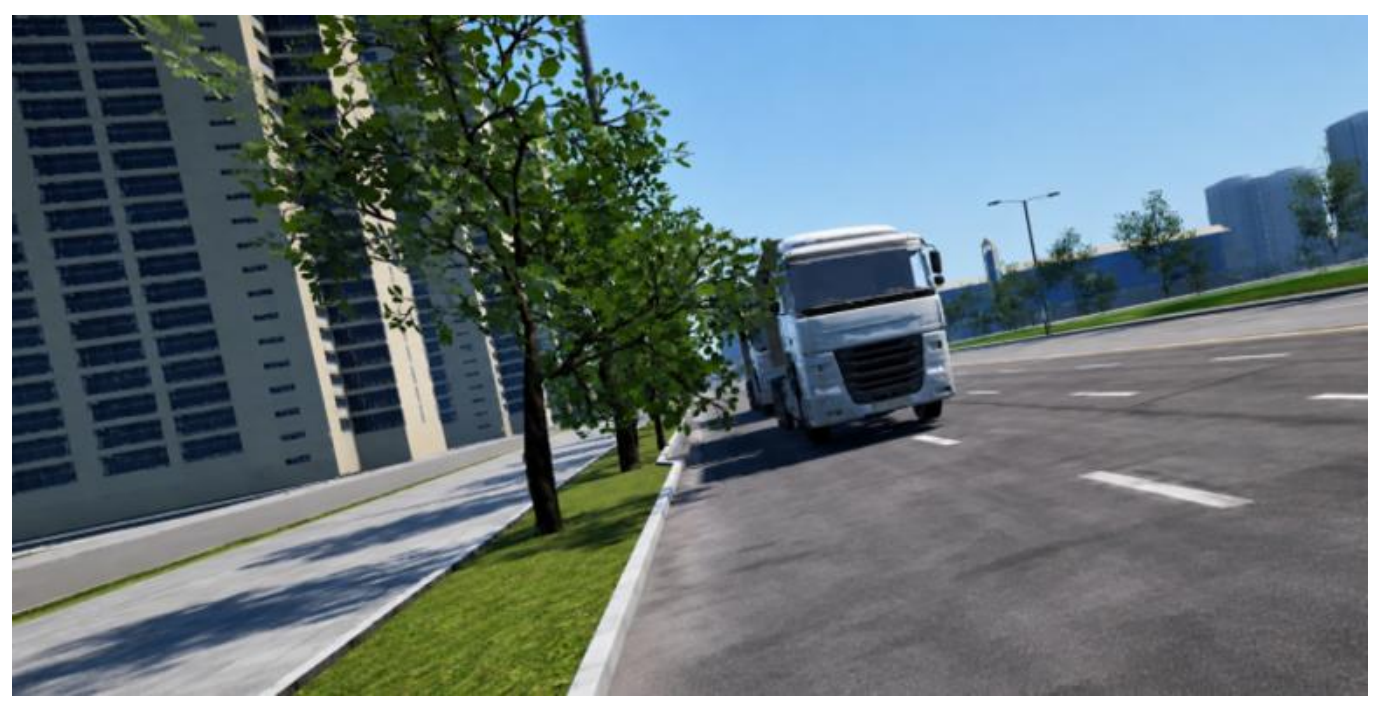

**Fig. 2.** Virtual reality experimental platform used for the pedestrian-autonomous truck crossing experiment: (a) participant setup with head-mounted display and controller; (b) first-person view of the unsignalized multi-lane crossing scenario in the virtual environment.

## 3.2 Experiment design

*3.2.1 Experiments scenarios*

A typical unsignalized two-lane crossing scenario was adopted to construct the pedestrian-vehicle interaction environment, with the scene design informed by Chen et al. (2025). The road was configured as a one-way facility with two lanes, consisting of a near-side lane and a far-side lane. Participants started from the curb and crossed toward the opposite side of the road. This setting was selected because it allowed the study to examine direct interaction with the near-side vehicle, occlusion caused by the near-side vehicle, braking-related safety margins, and subsequent interaction with the far-side vehicle within one continuous crossing process.

As shown in Fig. 3, the scenario included two near-side vehicle configurations. In the AT condition, two ATs were placed in the near-side lane and acted as both the direct interaction vehicles and the occluding vehicles. In the baseline condition, two passenger AVs were placed in the near-side lane. The far-side vehicle was kept as a passenger AV in both configurations. The two configurations shared the same road layout, crossing path, vehicle-triggering logic, and far-side vehicle setting, allowing near-side vehicle type and the associated differences in vehicle size, occlusion, and braking efficiency to be compared under consistent scenario conditions.

In both configurations, the scenario included two vehicles in the near-side lane and one vehicle in the far-side lane. The two near-side vehicles were arranged to maintain visual occlusion during the pedestrian's first-lane crossing process. The initial distance between the pedestrian and the leading near-side vehicle was 40 m, and the following near-side vehicle was positioned 5 m behind the leading vehicle. The following vehicle maintained the same motion state as the leading vehicle, including speed and acceleration. All vehicles had an initial speed of 50 km/h. The near-side vehicles were triggered to decelerate when they were 23 m from the pedestrian and stopped before the crossing path. The far-side vehicle approached at the same initial speed, and its yielding or non-yielding strategy was manipulated as part of the risk formation experiment.

Fig. 3 summarizes the basic scene layout and manipulated factors used in the risk formation experiment. The specific factor levels and experimental condition combinations are described in Section 3.2.2.

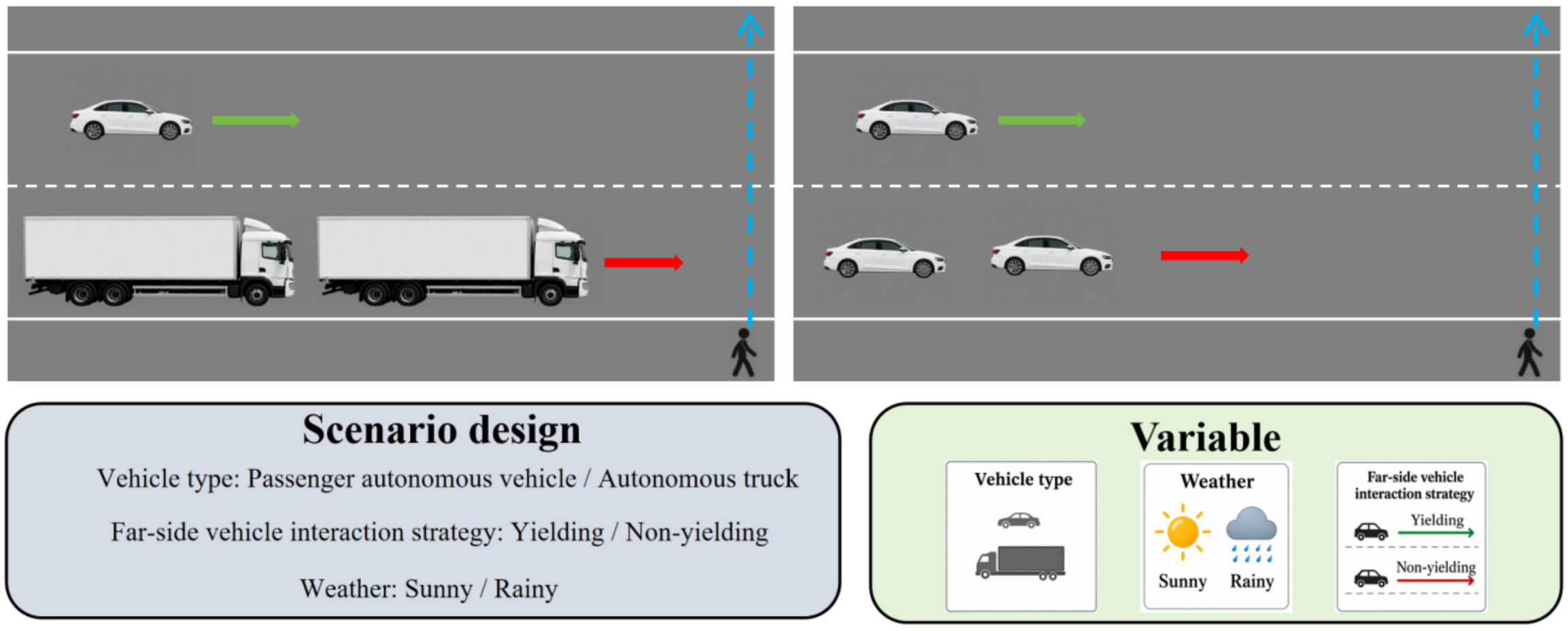

**Fig. 3.** Experimental scenario design.

*3.2.2 Parametric design*

The Stage 1 risk-formation component included three manipulated factors: near-side vehicle type, weather condition, and far-side vehicle interaction strategy. First, near-side vehicle type was manipulated to compare the AT and passenger-AV conditions. In the AT condition, the two near-side vehicles were ATs, with dimensions of 9.5 m in length, 4 m in width, and 4.1 m in height (General Administration of Quality Supervision, 2016). In the baseline condition, the near-side vehicles were passenger AVs, with dimensions of 5 m in length, 2 m in width, and 1.5 m in height (Tesla, 2024). This manipulation allowed the study to compare a high-occlusion AT condition with a passenger-AV condition in which the pedestrian's view of the far-side vehicle was less obstructed (Chen et al., 2025). Fig. 4 illustrates the relative difference in occlusion strength between the two near-side vehicle types, with the AT condition showing a more pronounced visibility restriction and the passenger-AV condition showing weaker occlusion of the far-side vehicle.

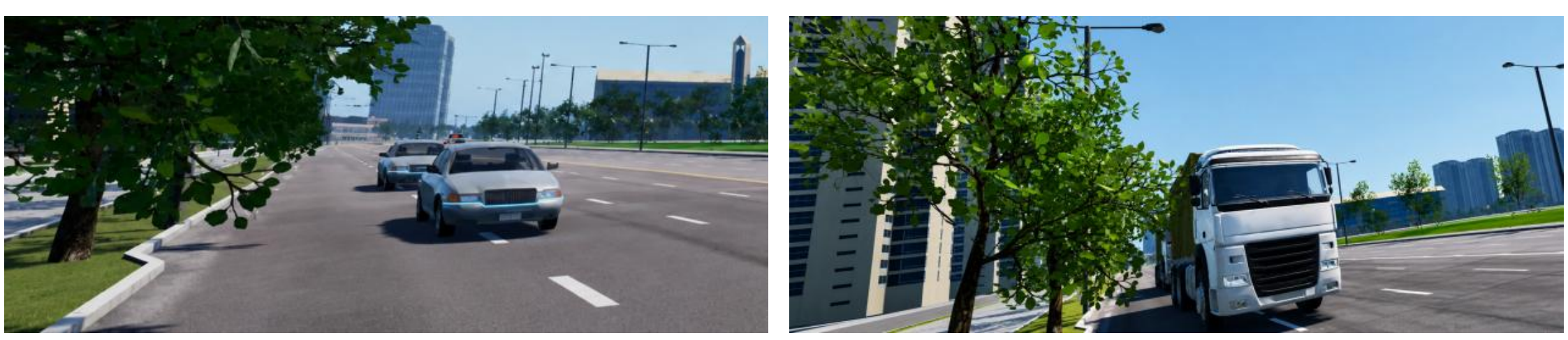

(a) Passenger-AV condition (b) AT condition

**Fig. 4.** Pedestrian-perspective comparison of occlusion in the AT and passenger-AV conditions.

Second, weather was manipulated using sunny and rainy conditions. Rainy weather was used to represent lower visual clarity and wet-road operation, which may increase crossing judgment difficulty and enlarge the braking disadvantage of trucks. Therefore, weather was treated as a factor affecting both pedestrian perception and vehicle deceleration capability.

Third, far-side vehicle interaction strategy was manipulated using yielding and non-yielding conditions. Under the yielding strategy, the far-side vehicle displayed the message "Vehicle yielding" on its roof and side and decelerated to stop when it approached within 10 m of the pedestrian. Under the non-yielding strategy, the far-side vehicle displayed the message "Vehicle non-yielding" and maintained its priority to proceed. These displays indicated the far-side vehicle's strategy and were distinct from the intervention measures evaluated in Stage 2. The interface forms of the two strategy indication displays are shown in Fig. 5.

(a) Yielding strategy

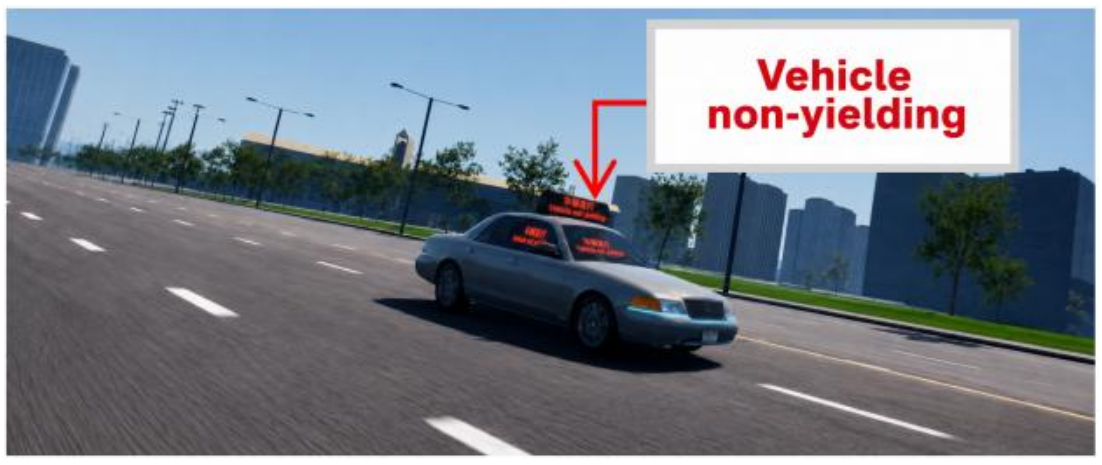

(b) Non-yielding strategy

**Fig. 5.** Strategy indication displays of the far-side AV.

Braking parameters were specified according to vehicle type and weather condition to represent differences in stopping capability. To maintain a consistent parameter basis, the deceleration settings for the passenger-vehicle and truck conditions at an initial speed of 50 km/h were calibrated with reference to the relevant Chinese vehicle safety standard (General Administration of Quality Supervision, 2017). The braking distances reported in Table 1 were then calculated from the assigned initial speed and calibrated deceleration values using the standard kinematic braking-distance relationship. The truck condition was assigned a lower deceleration and therefore a longer equivalent braking distance than the passenger-vehicle condition, while rainy conditions further reduced the calibrated deceleration to represent lower tire-road friction on wet pavement.

**Table 1. Equivalent braking distance settings calculated from calibrated deceleration parameters**

| Vehicle type | Weather | Braking distance (m) |
|---|---|---|
| Passenger autonomous vehicle | Sunny | 11.21 |
| Passenger autonomous vehicle | Rainy | 12.36 |
| Autonomous truck | Sunny | 18.01 |
| Autonomous truck | Rainy | 22.05 |

*3.2.3 Interventions*

During the experimental design, the representative high-risk scenario selected for intervention evaluation involved two ATs in the near-side lane, rainy weather, and a non-yielding far-side vehicle. Under this condition, pedestrians faced restricted visibility caused by the near-side truck, reduced safety margins under wet-road braking, and the need to identify the far-side risk source under occlusion. Therefore, the intervention measures were adapted to the risk characteristics of the AT scenario rather than directly reproducing conventional pedestrian-AV communication designs.

The intervention design followed three considerations. First, because the near-side AT created a large visual obstruction, the intervention should make pedestrians aware that the stopped truck could still create occlusion risk during subsequent crossing. Second, because the critical risk could arise from the far-side vehicle rather than the truck itself, the intervention should support recognition of the potential conflict area and risk location. Third, because pedestrians may not continuously attend to visual information, a non-visual warning channel was included to provide risk arousal under visually constrained conditions. Accordingly, three scenario-specific measures were designed and evaluated: environment-aware eHMI, projected eHMI, and auditory warning.

The environment-aware eHMI was designed as a vehicle-mounted occlusion warning on the near-side AT. The message "Caution blind spot" was displayed on the external body surface, as shown in Fig. 6(a). By attaching the warning to the truck that generated the occlusion, this intervention aimed to remind pedestrians that the visible stopping behavior of the near-side truck did not necessarily indicate that the subsequent lane was safe.

The projected eHMI was designed as a spatial warning linked to the far-side conflict area. A front-mounted projection module on the far-side vehicle displayed the message "Caution blind spot" on the road surface ahead of the vehicle, as shown in Fig. 6(b). Unlike the vehicle-mounted eHMI, this intervention placed the warning in the road environment to connect occlusion-related information with the potential conflict area and support pedestrians' spatial recognition of second-lane risk.

The auditory warning was designed as a non-visual risk-arousal measure. At the beginning of each trial, the formal interaction sequence was initiated when the participant entered a predefined trigger zone, which activated the vehicle movements and the corresponding experimental condition. In the auditory-warning condition, an alarm sound lasting 3 seconds was then delivered through the headset. This intervention provided an additional alert channel when visual information might be missed because of attention allocation to the truck, crossing path, or far-side vehicle. The sound duration, presentation channel, and triggering logic were standardized across auditory-warning trials.

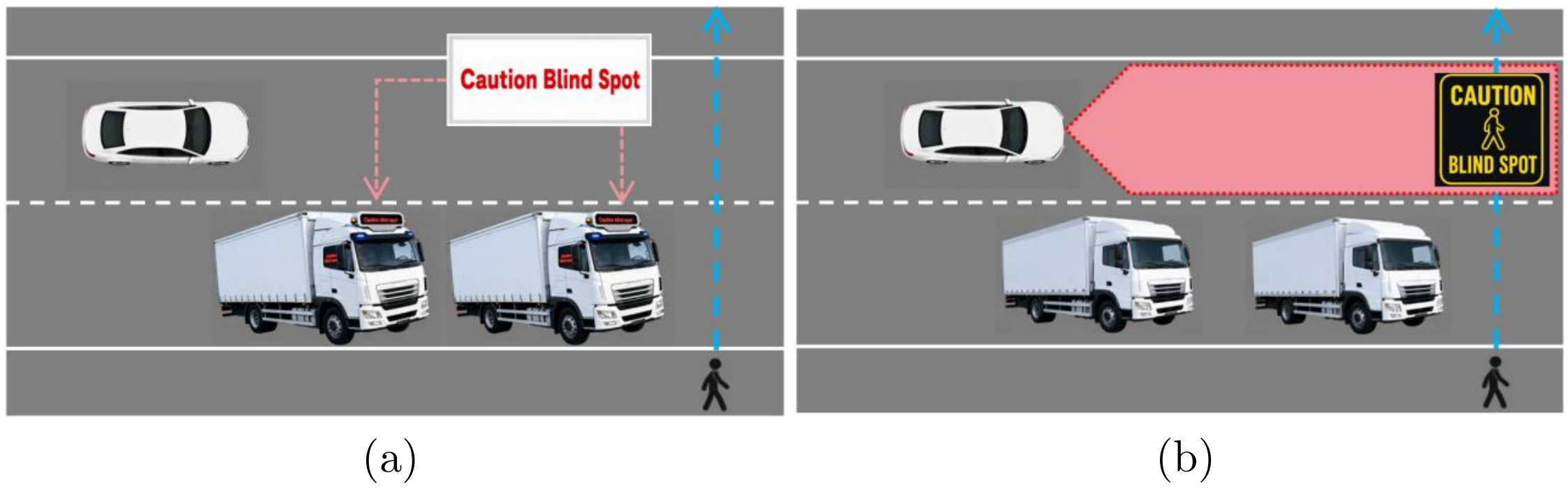

**Fig. 6.** Intervention measures: (a) Environment-aware eHMI; (b) Projected eHMI. Auditory warning was delivered through the headset and is not shown.

### 3.3 Experiment procedure

Prior to the study, the experimental protocol was reviewed and approved by the Science and Technology Ethics Committee of Tongji University. As illustrated in Fig. 7, the experimental procedure included informed consent, health-status screening, preparation, familiarization, and formal experiment. All trials were conducted on the same VR platform to ensure consistency and comparability across experimental conditions. Before execution, the participants received general information about the study, including the experimental tasks, use of VR equipment, potential discomfort associated with VR exposure, types of data to be collected, data confidentiality, and their right to withdraw from the experiment at any time. Written informed consent was obtained before any experimental screening or task was conducted.

After providing informed consent, participants underwent health-status screening and completed a pre-experiment questionnaire. The researchers then assisted them with headset setup and device calibration and introduced the controller functions, movement controls, and viewpoint adjustment. Participants assigned to auditory-warning conditions also confirmed that the warning sound was clearly audible through the headset.

During familiarization, participants practiced controller operation and the basic crossing task until they could navigate the virtual environment independently and confirmed that they understood the task requirements. They then completed the Virtual Reality Sickness Questionnaire (VRSQ; see Appendix A1), which assesses oculomotor discomfort and disorientation in virtual environments (Kim et al., 2018; Che et al., 2026). All participants reported no adverse symptoms and proceeded to the formal experiment.

The formal experiment consisted of 11 randomized trials for each participant, including eight risk-formation trials and three intervention-validation trials. The eight risk-formation trials corresponded to the combinations of near-side vehicle type, weather condition, and far-side vehicle interaction strategy. The three intervention-validation trials were conducted under the predefined representative high-risk scenario, namely ATs in the near-side lane, rainy weather, and a non-yielding far-side vehicle. A within-subject design was adopted, meaning that each participant experienced all experimental conditions. All 11 trials were randomized as a single trial set to reduce order effects and learning effects.

In each trial, participants completed the crossing task and immediately rated their perceived risk. A simulation-sickness check was then conducted after each trial. Participants reporting no symptoms proceeded to the next randomized trial, whereas the experiment would have been terminated if adverse symptoms were reported. All 54 participants reported no symptoms and completed all 11 trials, resulting in no exclusions or early terminations. After the final trial, participants completed the post-experiment questionnaire and received monetary compensation. During the formal experiment, the VR system automatically recorded participants' movement trajectories, crossing behavior, vehicle motion, and pedestrian-vehicle spatiotemporal relationships. Short breaks were arranged between trials when needed to reduce fatigue and repeated-exposure effects.

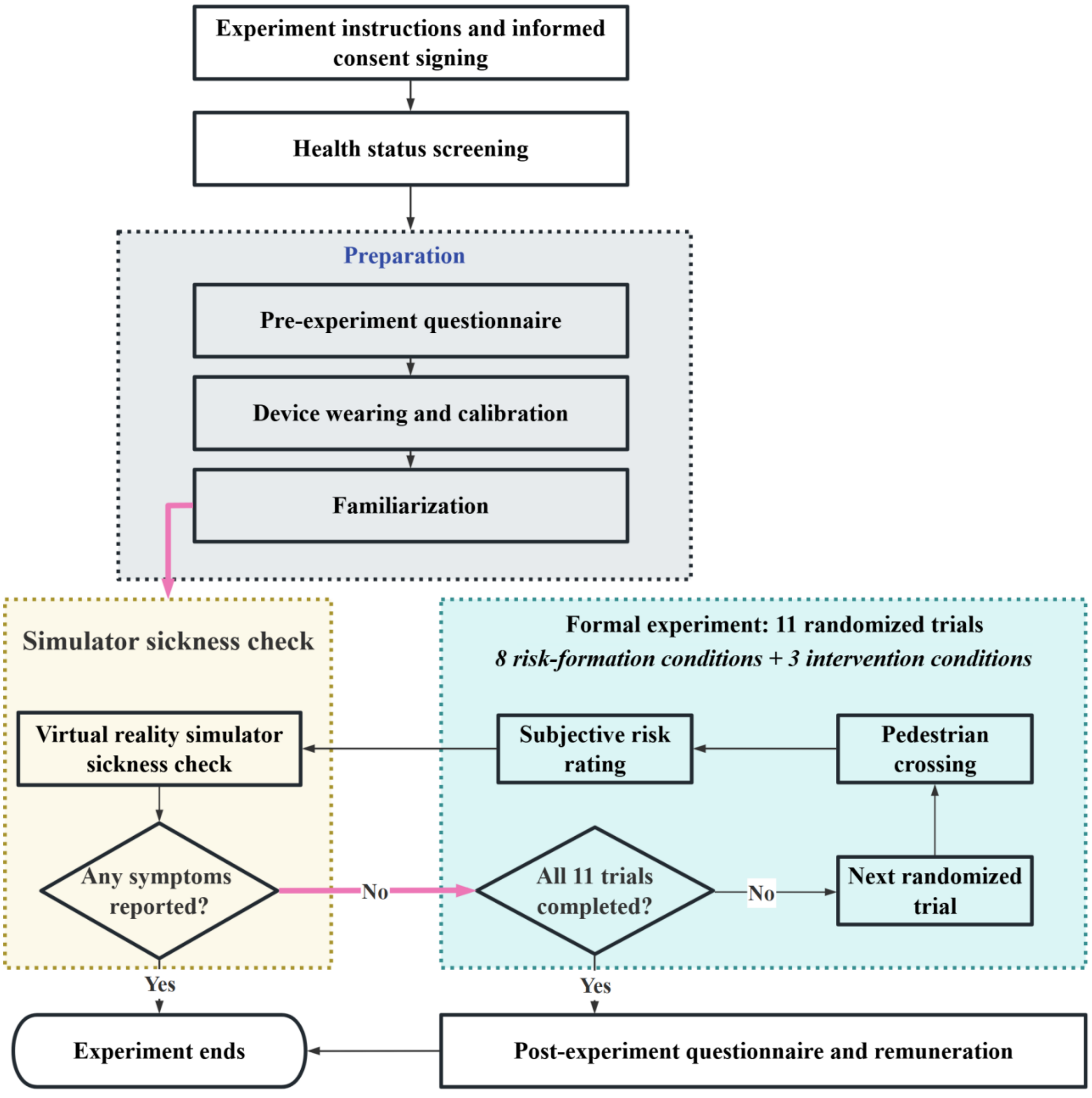

**Fig. 7.** Experimental procedure.

## 3.4 Experimental variables

*3.4.1 Demographic Variables*

A total of 54 participants were recruited through an electronic recruitment link distributed on the Tongji University campus, including 22 females and 32 males, ranging in age from 20 to 29 years. Their demographic information is presented in Table 2, including gender, age, education level and driving experience. These variables were used to describe the sample characteristics and to account for potential individual differences in driving-related experience and pedestrian crossing behavior.

**Table 2.** Demographic information

| Demographic variable | Category/Statistic | Value |
|---|---|---|
| Gender | Female | 22 (40.74%) |

| | Male | 32 (59.26%) |
|---|---|---|
| Age (years) | Mean (SD) | 24.04 (1.58) |
| | Range | 20–29 |
| Education level | Bachelor's degree or lower | 41 (75.93%) |
| | Master's degree or higher | 13 (24.07%) |
| Driving experience | 0 years | 8 (14.81%) |
| | Less than 1 year | 14 (25.93%) |
| | 1–3 years | 12 (22.22%) |
| | 3–More than 5 years | 15 (27.78%) |
| | >More than 5 years | 5 (9.26%) |

*3.4.2 Independent variables*

The independent variables were defined separately for the risk formation experiment and the intervention-effect experiment. In the risk formation experiment, the independent variables were near-side vehicle type, weather condition, and far-side vehicle interaction strategy. All three variables were binary categorical variables and were dummy-coded before being entered into the statistical models. Near-side vehicle type was coded as 0 = passenger AV and 1 = AT, weather condition as 0 = sunny and 1 = rainy, and far-side vehicle interaction strategy as 0 = yielding and 1 = non-yielding. These variables formed a 2 × 2 × 2 repeated-measures design, as shown in Table 3.

**Table 3.** Experimental conditions in the risk formation experiment

| **Scenario** | **near-side vehicle type** | **Weather condition** | **Far-side vehicle interaction strategy** |
|---|---|---|---|
| 1 | Passenger AVs | Sunny | Yielding |
| 2 | Passenger AVs | Sunny | Non-yielding |
| 3 | Passenger AVs | Rainy | Yielding |
| 4 | Passenger AVs | Rainy | Non-yielding |
| 5 | ATs | Sunny | Yielding |
| 6 | ATs | Sunny | Non-yielding |
| 7 | ATs | Rainy | Yielding |
| 8 | ATs | Rainy | Non-yielding |

The intervention-effect component focused on the AT-rainy-non-yielding condition selected as the representative high-risk scenario. The independent variable was intervention measure, with four types for analysis: no intervention, environment-aware eHMI, projected eHMI, and auditory warning. The no-intervention condition was also included in the risk formation component and served as the reference category for evaluating the three intervention measures. In the statistical models, the no-intervention condition was coded as 0, while each intervention measure was separately coded as 1 in its corresponding dummy variable. The intervention conditions are presented in Table 4.

**Table 4.** Intervention conditions in the intervention-effect experiment

| Scenario | Near-side vehicle type | Weather condition | Far-side vehicle interaction strategy | Intervention measure |
|---|---|---|---|---|
| 1 | ATs | Rainy | Non-yielding | Environment-aware eHMI |
| 2 | ATs | Rainy | Non-yielding | Projected eHMI |
| 3 | ATs | Rainy | Non-yielding | Auditory warning |
| 4 | ATs | Rainy | Non-yielding | None |

Note: Scenario 4 is the no-intervention baseline condition included among the risk-formation trials and served as the reference condition for the intervention-effect analysis. It was not repeated as a separate intervention-validation trial.

*3.4.3 Dependent variable*

The dependent variables were defined according to the objective of each analytical component. In the risk formation experiment, perceived risk and objective risk indicators were used to examine how AT-related factors shaped subjective risk judgment and objective interaction risk. In the intervention-effect experiment, perceived risk and objective risk indicators were retained, and behavioral indicators were additionally included to examine how pedestrians responded to different intervention measures.

**Subjective risk perception**. Perceived risk was used as a trial-level dependent variable to represent pedestrians' immediate subjective judgment of potential danger during each crossing interaction. Following previous VR-based pedestrian studies, subjective risk was collected immediately after each trial to capture participants' perceived risk in the just-experienced crossing scenario (Kwon et al., 2024). The rating was measured on a 10-point Likert scale, where 1 indicated extremely low risk and 10 indicated extremely high risk.

Given the sequential nature of the multi-lane crossing task, perceived risk was measured separately for the first-lane and second-lane interactions. This lane-specific measurement was adopted because participants first interacted with the near-side vehicle and then entered a subsequent interaction phase with the far-side vehicle under possible occlusion. The descriptions of the perceived risk variables are provided in Table 5. Higher scores indicated higher perceived risk in the corresponding lane-level interaction.

**Table 5.** Description of perceived risk variables used in the experiment

| Variable | Description | Measurement timing | Scale |
|---|---|---|---|
| First-lane perceived risk | Participants rated the perceived danger of their interaction with the vehicle in the first lane. | After each trial | 1-10 |
| Second-lane perceived risk | Participants rated the perceived danger of their interaction with the vehicle in the second lane. | After each trial | 1-10 |

Note: Higher scores indicate higher perceived risk. The two ratings were collected after each trial.

**Objective risk indicators.** Objective risk was quantified using lane-specific surrogate safety indicators. The conflict zone was defined as the spatial intersection between the trajectories of the pedestrian and the vehicle. In the multi-lane crossing scenario, the conflict zone was divided into the first-lane conflict zone and the second-lane conflict zone. Because the risk sources differed between the two lanes, different surrogate safety indicators were used.

For the first lane, deceleration to safety time (DST) was adopted as the core objective risk measure (Hupfer, 1997). DST is defined as the ratio between the minimum deceleration required for the vehicle to avoid a collision and the vehicle's actual average deceleration. Because DST is sensitive to vehicle dynamics and to changes in braking efficiency caused by weather conditions, it is suitable for capturing the safety margin during the direct interaction between pedestrians and the near-side vehicle. Its calculation is given in Eq. (1).

$$DST = \frac{v_0}{a_{act}(t_1 - t_0)} \tag{1}$$

where $t_0$ denotes the predicted arrival time of the vehicle at the conflict zone in the first lane if it continues at its current speed without decelerating; $t_1$ denotes the time at which the pedestrian leaves the first-lane conflict zone; $v_0$ denotes the initial approach speed of the vehicle before deceleration is triggered; and $a_{act}$ denotes the vehicle's actual average deceleration. DST is dimensionless rather than being expressed in seconds. If DST is negative, the vehicle would still reach the conflict point only after the pedestrian has left the first-lane conflict zone, even without additional braking. This indicates that a certain temporal safety margin remains between the pedestrian and the vehicle, and that the objective interaction risk in the first lane is relatively low. As DST approaches zero or increases further, the temporal separation between the vehicle and the pedestrian in the conflict zone becomes smaller, the temporal safety redundancy is reduced, and the objective interaction risk correspondingly increases.

For the second lane, improved time advantage (ITADV) was adopted as the core objective risk measure (Sheng et al., 2025). ITADV describes the evolution of the time advantage between the pedestrian and the vehicle near the potential conflict zone, given in Eqs. (2)-(4).

$$TADV = \frac{d_{veh}}{v_{veh}} - \frac{d_{ped}}{v_{ped}} \tag{2}$$

$$T_2 = \max\{\frac{d_{veh}}{v_{veh}}, \frac{d_{ped}}{v_{ped}}\} \tag{3}$$

$$ITADV = \min\{TADV \mid 0 < T_2 < T_2^*\} \tag{4}$$

where $d_{veh}$ and $v_{veh}$ denote the distance and speed of the vehicle at a given moment, respectively, while $d_{ped}$ and $v_{ped}$ denote the pedestrian's distance to the conflict zone and walking speed at that moment. $T_2$ denotes the time required for the slower of the two, namely the pedestrian or the vehicle, to reach the conflict zone, and $T_2^*$ is generally set to 2.5 s (Sheng et al., 2025).

ITADV measures the pedestrian's temporal advantage over the vehicle during their approach to the second-lane conflict zone, while incorporating a minimum safety-time threshold that represents the temporal separation adopted for evaluating crossing safety. ITADV is expressed in seconds because it is derived from the difference between the estimated arrival times of the pedestrian and the vehicle. A larger ITADV value indicates greater temporal separation and a larger safety margin, whereas a smaller value indicates reduced temporal redundancy and higher interaction risk. This indicator is particularly suitable for the second lane because, under occlusion created by the near-side vehicle, safety depends on whether the pedestrian can recognize the far-side risk source in time and retain sufficient temporal advantage before entering the potential conflict area.

**Behavioral indicators.** Behavioral indicators were included in the intervention-effect experiment to clarify how pedestrians responded to different forms of risk information. The indicators included waiting time, crossing time, and vehicle observation time. Waiting time refers to the time a pedestrian remains stationary before entering a given lane, reflecting observation, hesitation, and timing selection before crossing. Crossing time refers to the duration of traversing a given lane, reflecting exposure duration and movement rhythm. Vehicle observation time refers to the duration for which the pedestrian's gaze was directed toward the relevant vehicle, reflecting attention allocation and visual information search during the interaction. This indicator was derived from eye-tracking data recorded by the VR headset. For each recorded frame, the gaze-point coordinates were compared with a dynamic area of interest defined by the coordinates covered by the relevant vehicle body in the participant's visual field. A frame was classified as vehicle observation when the gaze point fell within the vehicle-body area, and vehicle observation time was calculated by summing the duration of all qualifying frames.

These behavioral indicators supplemented the interpretation of perceived risk and objective risk outcomes. They helped determine whether an intervention mainly increased risk awareness, changed the crossing process, or generated improvements in objective safety margins.

### 3.5 Statistical modeling

Descriptive statistics, linear mixed models (LMMs), and mediation analysis were used to examine pedestrian-AT interaction risk and intervention effects. Descriptive statistics summarized perceived risk, behavioral indicators, and objective risk measures across lanes and experimental conditions. LMMs estimated the effects of experimental factors while accounting for repeated observations from the same participants, and mediation analysis examined whether the effect of the AT condition on second-lane objective risk was transmitted through second-lane entry timing or remained as an occlusion-related effect.

#### *3.5.1 Descriptive statistics*

Descriptive statistics were used to summarize the distributional characteristics of perceived risk, behavioral indicators, and objective risk indicators. For each lane and experimental condition, the mean, standard deviation, minimum, and maximum values were calculated to provide an initial overview of pedestrians' behavioral responses and risk levels before model-based analysis.

The descriptive results also supported the subsequent model-based analyses. By comparing the variation patterns of each indicator across near-side vehicle type, weather condition, far-side vehicle interaction strategy, and intervention measure, it was possible to preliminarily identify whether systematic differences existed across experimental conditions. In addition, comparing first-lane and second-lane indicators helped justify the lane-specific risk assessment adopted in this study.

#### *3.5.2 Linear mixed models*

Because the experiment adopted a within-subject design, each participant contributed multiple observations under different experimental conditions. This repeated-measures structure creates within-subject correlation and violates the independence assumption of conventional linear regression. Therefore, LMMs with participant-level random intercepts were used to account for individual baseline differences in perceived risk, behavioral performance, and objective risk level (Laird and Ware, 1982; Subramanian et al., 2024). The general form of the LMM is given as:

$$Y_{ij} = \beta_0 + \boldsymbol{\beta}^T \mathbf{X}_{ij} + u_i + \varepsilon_{ij} \tag{5}$$

where $Y_{ij}$ denotes the dependent variable for participant $i$ in trial $j$, $\mathbf{X}_{ij}$ represents the vector of fixed-effect predictors, $\boldsymbol{\beta}$ denotes the corresponding fixed-effect coefficients, $u_i$ is the participant-level random intercept, and $\varepsilon_{ij}$ is the residual error. The random intercept captures stable between-participant differences, such as individual differences in risk preference, crossing style, and traffic experience.

For the risk formation experiment, the fixed effects included near-side vehicle type, weather condition, far-side vehicle interaction strategy, and the interaction between vehicle type and weather condition. These models were used to examine how AT-related characteristics and complex traffic conditions shaped perceived risk and objective interaction risk. For the intervention-effect experiment, auditory warning, projected eHMI, and environment-aware eHMI were entered as intervention indicators, with the no-intervention high-risk condition used as the reference. These models were used to evaluate whether the interventions produced significant changes in perceived risk, behavioral indicators, and objective safety measures.

#### *3.5.3 Mediation analysis*

A mediation analysis was conducted to examine how the near-side AT condition influenced second-lane objective risk (Zhang et al., 2025). As shown in Fig. 8, the analysis separated the total AT effect into a timing-mediated indirect pathway through second-lane entry time and a remaining direct occlusion pathway. Because each participant contributed repeated observations, the three equations were estimated as linear mixed-effects models with participant-specific random intercepts.

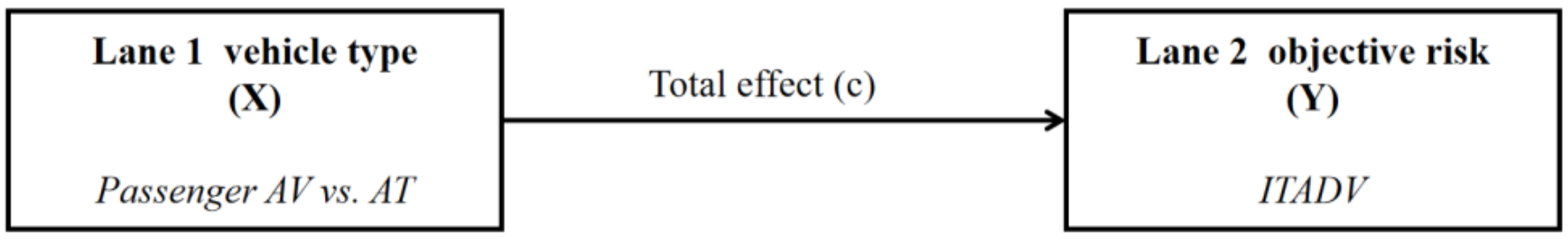

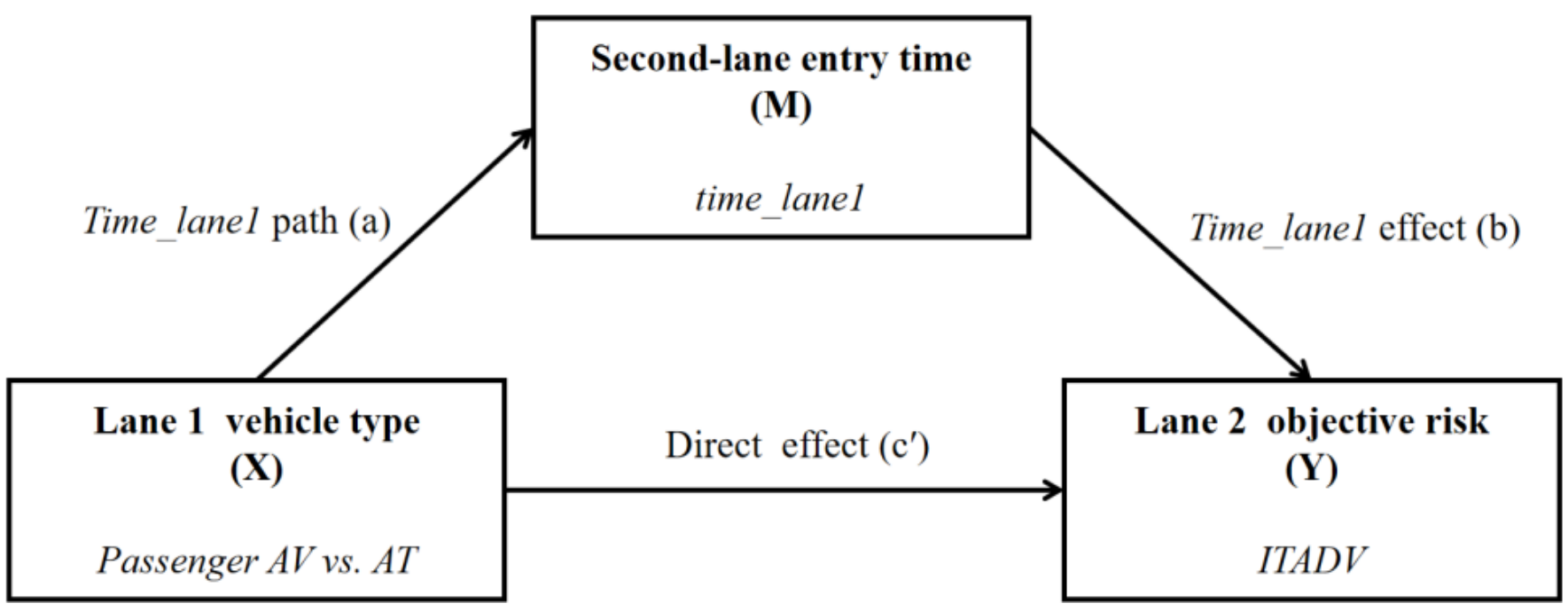

**Fig. 8**. Analytical framework of mediation analysis.

Near-side vehicle type was denoted by $X_{ij}$, where $X_{ij}=0$ represented the passenger-AV condition and $X_{ij}=1$ represented the AT condition. Second-lane entry time was denoted by $T_{entry,ij}$ and defined as the interval from vehicle triggering to the pedestrian reaching the second-lane conflict zone. The outcome $Y_{ij}$ was second-lane ITADV; lower ITADV indicates a smaller temporal advantage and higher objective risk. Subscripts $i$ and $j$ identify participant $i$ and experimental condition $j$, respectively. Weather, far-side interaction strategy, gender, driving experience and education were included as covariates and represented collectively by $Z_{ij}$.

First, the total effect of the AT condition on ITADV was estimated as follows:

$$Y_{ij}=\alpha_0+cX_{ij}+\boldsymbol{\theta}^T\mathbf{Z}_{ij}+u_{0i}^{(1)}+\varepsilon_{ij}^{(1)} \tag{6}$$

where $\alpha_0$ is the fixed intercept; $c$ is the total AT effect on ITADV; $\theta$ is the corresponding coefficient vector; $\boldsymbol{\theta}^T\mathbf{Z}_{ij}$ is the combined covariate contribution; $u_{0i}^{(1)}$ is the participant-specific random intercept; and $\varepsilon_{ij}^{(1)}$ is the observation-level residual in the total-effect model.

Second, the effect of the AT condition on $T_{entry,ij}$ was estimated as follows:

$$T_{entry,ij}=\beta_0+aX_{ij}+\Phi^T\mathbf{Z}_{ij}+u_{0i}^{(2)}+\varepsilon_{ij}^{(2)} \tag{7}$$

where $\beta_0$ is the fixed intercept; $a$ is the effect of the AT condition on entry time; $\Phi$ is the covariate-coefficient vector; $\Phi^T\mathbf{Z}_{ij}$ is the combined covariate contribution; $u_{0i}^{(2)}$ is the participant-specific random intercept; and $\varepsilon_{ij}^{(2)}$ is the residual in the mediator model. A positive $a$ indicates later entry under the AT condition.

Third, vehicle type and entry time were entered simultaneously into the ITADV model:

$$Y_{ij} = \gamma_0 + c' X_{ij} + bT_{entry,ij} + \Psi^T \mathbf{Z}_{ij} + u_{0i}^{(3)} + \varepsilon_{ij}^{(3)} \tag{8}$$

In Eq. (8), $\gamma_0$ is the fixed intercept; $c'$ is the direct AT effect after entry time is controlled for; $b$ is the association between entry time and ITADV; $T_{entry,ij}$ is the mediator; $\Psi$ is the covariate-coefficient vector; $\Psi^T \mathbf{Z}_{ij}$ is the combined covariate contribution; $u_{0i}^{(3)}$ is the participant-specific random intercept; and $\varepsilon_{ij}^{(3)}$ is the residual. In the present experiment, $c'$ was interpreted as the direct occlusion effect. The random intercepts and residuals were assumed to be normally distributed:

$$u_{0i}^{(k)} \sim N(0, \sigma_{u,k}^2),\ \varepsilon_{ij}^{(k)} \sim N(0, \sigma_{\varepsilon,k}^2),\ k = 1,2,3 \tag{9}$$

where $k$ identifies the three equations; $u_{0i}^{(k)}$ is the participant-specific random intercept in equation $k$; $\sigma_{u,k}^2$ is its between-participant variance; $\varepsilon_{ij}^{(k)}$ is the residual in equation $k$; and $\sigma_{\varepsilon,k}^2$ is the corresponding within-participant residual variance.

The timing-mediated indirect effect was calculated as $ab$. A negative $ab$ indicates that the AT condition increased objective risk through the timing pathway. The total effect was decomposed as:

$$c = c' + ab \tag{10}$$

where $c$ is the total AT effect, $c'$ is the direct occlusion effect. The descriptive proportion mediated by entry time was calculated as:

$$P_{med} = ab/c \times 100\% \tag{11}$$

This proportion was interpreted only when $c$, $c'$ and $ab$ had the same direction and $c$ was nonzero. Because the sampling distribution of $ab$ may be asymmetric, inference was based on a participant-level cluster Bootstrap. Participants were sampled with replacement, and all repeated observations from each sampled participant were retained together. The indirect effect in each replication was calculated as:

$$\widehat{IE}^{(r)} = \hat{a}^{(r)}\hat{b}^{(r)}, r = 1,2,3, \dots, B \tag{12}$$

where $\widehat{IE}^{(r)}$ is the estimated indirect effect in Bootstrap replication $r$; $\hat{a}^{(r)}$ and $\hat{b}^{(r)}$ are the estimated pathway coefficients in that replication; $r$ indexes the replication; and $B$ is the requested number of replications, set to $B = 5{,}000$.

Bias-corrected 95% Bootstrap confidence intervals were used. An effect was considered statistically significant when its interval did not include zero; significant indirect and direct effects indicated partial mediation.

# 4. Results

## 4.1 Descriptive analysis

### *4.1.1 Descriptive results of the risk formation experiment*

Table 6 presents the descriptive statistics for the risk formation experiment, while the condition-wise descriptive statistics across the eight experimental conditions are provided in Appendix A2. The results provide an initial overview of pedestrians' perceived risk and lane-specific objective safety indicators in the multi-lane crossing scenario.

Several preliminary patterns can be observed. First, the mean DST was negative (Mean = -0.18), suggesting that, on average, a certain temporal safety margin remained in the first-lane interaction. However, the large standard deviation (SD = 1.11) and wide value range indicated substantial heterogeneity in first-lane safety margins across experimental conditions. Second, the mean ITADV was 1.43, but its relatively large dispersion (SD = 3.65) suggested that the time advantage between pedestrians and the far-side vehicle varied considerably in the second-lane interaction. This indicates that second-lane safety was not stable across conditions, even though pedestrians generally retained some time advantage before entering the second-lane conflict area. Third, perceived risk showed a lane-related difference. The mean perceived risk was higher in the second lane (Mean = 7.51) than in the first lane (Mean = 6.51), suggesting that pedestrians subjectively perceived the second-lane interaction as more dangerous.

Overall, the descriptive results suggest that pedestrian-vehicle interaction risk in the multi-lane scenario was phase-dependent. The first-lane interaction was mainly reflected in variations in braking-related safety margins, whereas the second-lane interaction showed greater instability in time advantage and higher perceived risk. These preliminary patterns provide the basis for the subsequent model-based analysis.

**Table 6.** Descriptive statistics for the risk formation experiment

| Variable | Mean | SD | Min | Max |
|---|---|---|---|---|
| DST | -0.18 | 1.11 | -13.73 | 0.14 |
| ITADV | 1.43 | 3.65 | 0.01 | 8.81 |
| Risk perception (Lane 1) | 6.51 | 1.75 | 3.00 | 10.00 |
| Risk perception (Lane 2) | 7.51 | 0.94 | 5.00 | 10.00 |

### *4.1.2 Descriptive results of the intervention-effect experiment*

Descriptive statistics for the intervention-effect experiment are presented in Table 7, and the condition-wise descriptive results for the four intervention conditions are provided in Appendix A3. These results provide an initial overview of pedestrian responses under the selected high-risk AT scenario.

Three preliminary patterns can be observed. First, the mean DST was close to zero (Mean = -0.01), while the mean ITADV was relatively low (Mean = 1.19). This suggests that the intervention-effect experiment was conducted under a constrained safety condition, especially with respect to the time advantage available in the second-lane conflict area. Second, behavioral indicators showed clear differences between the two crossing lanes. Waiting time and vehicle observation time were higher in the first lane (Mean = 8.66 and Mean = 4.85) than in the second lane (Mean = 2.70 and Mean = 0.64). By contrast, crossing time was relatively similar between the first lane (Mean = 2.80) and the second lane (Mean = 2.52). This indicates that behavioral differences were mainly reflected in observation and timing decisions rather than in walking execution. Third, perceived risk remained high in both lanes, with mean values of 8.93 in the first lane and 8.11 in the second lane, suggesting that participants generally recognized the selected scenario as risky.

**Table 7.** Descriptive statistics for the intervention-effect experiment

| Variable | Mean | SD | Min | Max |
|---|---|---|---|---|
| DST | -0.01 | 0.36 | -2.99 | 0.14 |
| ITADV | 1.19 | 1.97 | 0.01 | 5.652 |
| Waiting time (Lane 1) | 8.66 | 2.42 | 3.37 | 15.75 |
| Waiting time (Lane 2) | 2.70 | 1.22 | 1.55 | 6.45 |
| Crossing time (Lane 1) | 2.80 | 1.11 | 1.64 | 5.84 |
| Crossing time (Lane 2) | 2.52 | 0.44 | 2.07 | 6.28 |
| Vehicle observation time (Lane 1) | 4.85 | 2.36 | 0.02 | 11.50 |
| Vehicle observation time (Lane 2) | 0.64 | 0.81 | 0.02 | 4.95 |
| Risk perception (Lane 1) | 8.93 | 0.77 | 7.00 | 10.00 |
| Risk perception (Lane 2) | 8.11 | 0.76 | 6.00 | 10.00 |

The descriptive results suggest that the two experimental parts captured different levels of interaction risk. In the risk formation experiment, objective safety indicators showed considerable variation across vehicle, weather, and interaction-strategy conditions. In contrast, the intervention-effect experiment focused on a more constrained high-risk scenario, characterized by DST values close to the critical boundary, lower ITADV, and high perceived risk in both lanes. This contrast supports the subsequent analysis: the first part identifies how pedestrian-ATs interaction risk is formed, whereas the second part examines whether targeted interventions can improve behavior and safety under a more critical interaction context.

**4.2 Risk formation experiment**

The risk formation experiment examined how ATs-related factors shaped pedestrian interaction risk under different vehicle combinations of near-side vehicle type, weather condition, and far-side vehicle interaction strategy. This section first reports perceived and objective risk during the direct first-lane interaction. It then examines perceived risk during the subsequent second-lane interaction, followed by an analysis of second-lane objective risk. In particular, ITADV and mediation analysis were used to distinguish the effect transmitted through second-lane entry time from the remaining occlusion-related effect.

*4.2.1 Risk perception in the first lane*

The LMM results for first-lane perceived risk are presented in Table 8. Vehicle type had a significant positive effect on perceived risk ($\beta = 3.046$, $p < 0.001$). Compared with the passenger AV condition, the ATs condition led to substantially higher perceived risk in the first lane. This indicates that pedestrians were sensitive to truck-related risk attributes during direct interaction, such as larger body size, stronger visual pressure, and potentially more severe conflict consequences.

**Table 8.** LMM results for first-lane perceived risk

| Variable | Coefficient | SE | *z*-value | *p*-value |
|---|---|---|---|---|
| | Fixed effects | | | |
| Vehicle type | 3.046 | 0.092 | 33.28 | <0.001 |
| Weather condition | 1.000 | 0.092 | 10.93 | <0.001 |
| | Random effects | | | |
| Variance of random intercept | 0.050 | 0.021 | | |

SE = standard error.

Weather condition also showed a significant positive effect ($\beta = 1.000$, $p < 0.001$), indicating that rainy conditions further increased pedestrians' perceived risk in the first lane. This suggests that pedestrians did not treat weather merely as a visual background, but incorporated adverse environmental conditions into their risk judgment during interaction with the near-side vehicle.

The participant-level random-intercept variance was 0.050, indicating that baseline perceived risk differed across participants and supporting the use of a mixed-effects model for repeated observations.

*4.2.2 Objective risk in the first lane*

The LMM results for first-lane objective risk, measured by DST, are presented in Table 9. Because a lower DST value indicates a larger temporal safety margin, negative coefficients should be interpreted as a reduction in objective interaction risk.

**Table 9.** LMM results for first-lane objective risk measured by DST

| Variable | Coefficient | SE | *z*-value | *p*-value |
|---|---|---|---|---|
| | Fixed effects | | | |
| Vehicle type | -0.290 | 0.137 | -2.12 | 0.034 |
| Vehicle type × Weather | 0.383 | 0.194 | 1.98 | 0.048 |
| | Random effects | | | |
| Variance of random intercept | 0.168 | 0.058 | | |

SE = standard error.

As shown in Table 9, vehicle type had a significant negative effect on DST ($\beta = -0.290$, $p < 0.05$). Compared with the passenger AV condition, the ATs condition was associated with a lower DST value, indicating a larger temporal safety margin in the first-lane interaction. This result does not imply that ATs were inherently safer. Rather, it suggests a possible risk compensation effect: pedestrians may have responded to the larger body size and stronger visual pressure of ATs by choosing more conservative crossing timing, thereby maintaining greater safety redundancy during direct interaction.

However, this safety-margin pattern changed under rainy conditions. Although weather condition alone did not significantly affect DST in the passenger AV condition, the interaction between vehicle type and weather condition was significant and positive ($\beta = 0.383$, $p < 0.05$), showing that the lower DST associated with ATs was weakened when the weather condition became adverse. This finding suggests that pedestrian caution may partly compensate for the risk posed by ATs under normal conditions, but such compensation becomes less effective when rainy weather further reduces braking efficiency. In this sense, weather condition did not simply act as a background factor; it altered how truck-related dynamic constraints were reflected in the objective risk outcome.

The random-intercept variance was 0.168, indicating that baseline objective risk levels varied across participants and supporting the use of a mixed-effects model for repeated observations.

Viewed across the two first-lane models, perceived risk and objective risk showed a meaningful divergence. Pedestrians' perceived risk increased directly under both ATs and rainy conditions, suggesting that their subjective judgments were mainly driven by salient external cues, such as vehicle size, visual pressure, and adverse weather. However, the objective DST results showed a different pattern. The ATs condition was associated with a larger temporal safety margin, which points to a possible risk compensation response, whereas this margin was reduced when rainy conditions constrained braking efficiency. More importantly, the braking-efficiency interaction affected objective risk but did not appear in the perceived-risk model. This suggests that pedestrians could recognize the apparent danger of ATs and rain, but may not fully internalize how the combination of truck type and reduced braking efficiency changes the actual safety margin. Thus, the first-lane results reveal a gap between subjective risk awareness and objective dynamic risk.

*4.2.3 Risk perception in the second lane*

The LMM results for perceived risk in the subsequent interaction phase are presented in Table 10. As shown in Table 10, vehicle type had a significant positive effect on perceived risk in the subsequent interaction phase ($\beta = 1.009$, $p < 0.001$). Compared with the passenger AVs condition, the ATs condition led to higher perceived risk, indicating that pedestrians remained sensitive to truck-related risk even after the direct interaction with the near-side vehicle.

**Table 10.** LMM results for second-lane perceived risk

| **Variable** | **Coefficient** | **SE** | ***z*-value** | ***p*-value** |
|---|---|---|---|---|
| | Fixed effects | | | |
| Vehicle type | 1.009 | 0.092 | 10.93 | <0.001 |
| Weather condition | 0.370 | 0.092 | 4.01 | <0.001 |
| Interaction strategy | 0.778 | 0.065 | 11.91 | <0.001 |
| | Random effects | | | |
| Variance of random intercept | 0.011 | 0.014 | | |

SE = standard error.

Weather condition also had a significant positive effect ($\beta = 0.370$, $p < 0.001$), suggesting that rainy conditions further increased pedestrians' perceived risk in the subsequent interaction phase. In addition, far-side vehicle interaction strategy showed a significant positive effect ($\beta = 0.778$, $p < 0.001$), indicating that pedestrians' risk perception was also shaped by the behavioral response of the far-side vehicle.

The random-intercept variance was relatively small (0.011). However, the likelihood-ratio test indicated that the random-intercept model did not provide a statistically significant improvement in fit over the corresponding linear model ($\bar{\chi}^2(1) = 0.81$, $p = 0.184$). Thus, the results provided limited evidence of stable between-participant differences in baseline second-lane perceived risk. Nevertheless, the random intercept was retained to account for the repeated-measures structure and to maintain a consistent model specification across the risk-formation outcomes.

Age, driving-experience, and education level were included as covariates in all three models. None of these variables reached the conventional $p < 0.05$ significance threshold.

*4.2.4 Objective risk in the second lane*

ITADV was used to evaluate objective risk during pedestrians' subsequent interaction with the far-side vehicle. A mediation analysis was conducted to distinguish the indirect effect transmitted through second-lane entry time from the remaining direct effect of the AT condition. The models controlled for weather condition, far-side vehicle interaction strategy, gender, age, driving experience, and education level. Statistical inference for the mediation effects was based on an education-stratified, participant-level cluster bootstrap. The mediation results are presented in Table 11, and the complete estimates for the three component models are reported in Appendix A4-A7.

**Table 11.** Mediation effects on second-lane ITADV

| **Effect** | **Coefficient** | **Bootstrap SE** | **z-value** | **p-value** | **95% bias-corrected bootstrap CI** |
|---|---|---|---|---|---|
| AT → $T_{entry}$ | 1.004 | 0.126 | 7.95 | <0.001 | [0.756, 1.251] |
| $T_{entry}$ → ITADV | -0.403 | 0.082 | -4.91 | <0.001 | [-0.565, -0.242] |
| Indirect effect ($ab$) | -0.405 | 0.095 | -4.25 | <0.001 | [-0.592, -0.218] |
| Direct occlusion effect ($c'$) | -1.089 | 0.146 | -7.48 | <0.001 | [-1.375, -0.804] |
| Total effect ($c$) | -1.494 | 0.171 | -8.74 | <0.001 | [-1.829, -1.159] |
| Mediated proportion | 27.10% | — | — | — | — |
| Direct-effect proportion | 72.90% | — | — | — | — |

Note: Bootstrap estimates were based on 4,981 complete replications out of the 5,000 requested; model parameters could not be estimated for 19 replications.

As shown in Table 11, the AT condition had a significant negative total effect on ITADV ($c = -1.494$, 95% bias-corrected bootstrap CI [-1.829, -1.159]). Compared with the passenger-AV condition, the AT condition therefore reduced pedestrians' temporal advantage relative to the far-side vehicle and increased second-lane objective risk.,

In the mediator model specified in Eq. (7), the AT condition increased $T_{entry}$ by approximately 1.004 s ($a = 1.004$, 95% bias-corrected bootstrap CI [0.756, 1.251]). Thus, pedestrians entered the second-lane interaction phase later under the AT condition than under the passenger-AV condition.

In the outcome model specified in Eq. (8), second-lane entry time was negatively associated with ITADV after near-side vehicle type and the covariates were controlled for ($b = -0.403$, 95% bias-corrected bootstrap CI [-0.565, -0.242]). Each additional second before entering the second-lane interaction phase was associated with an approximately 0.403 s reduction in ITADV. The two component coefficients, $a$ and $b$, are reported in the first two rows of Table 11.

The mediated indirect effect was significant ( $ab = -0.405$ , 95% bias-corrected bootstrap CI [-0.592, -0.218]). This result indicates that the AT condition increased second-lane objective risk partly through changes in second-lane entry time. According to Eq. (11), the ratio of the indirect effect $ab$, to the total effect $c$, was 27.10%, as reported in Table 11.

After $T_{entry}$ was included in the outcome model, the direct effect of the AT condition remained significant ( $c' = -1.089$, 95% bias-corrected bootstrap CI [-1.375, -0.804]). In the present experimental context, this remaining effect was interpreted as the direct occlusion effect and accounted for 72.90% of the total effect. This finding indicates that the large AT body continued to restrict pedestrians' recognition of the far-side vehicle after differences in entry timing were controlled for.

Because both the indirect effect $ab$, and the direct occlusion effect $c'$, were statistically significant, second-lane entry time partially mediated the relationship between the AT condition and ITADV. The AT condition therefore increased second-lane objective risk through both changes in entry timing and the remaining visibility restriction associated with the larger truck body.

Complete estimates for the total-effect, mediator, and direct-effect models are reported in Appendices A4-A6, respectively, with random-effects and model-fit statistics provided in Appendix A7. In the mediator model, rainy weather was associated with a 0.272-s increase in second-lane entry time ( $\beta = 0.272$ , $p < 0.05$ ), whereas male participants entered the second-lane conflict zone approximately 1.262 s earlier than female participants ( $\beta = -1.262$ , $p < 0.05$ ). In the total-effect model and the direct-effect model, none of the covariates reached the $p<0.05$ significance threshold.

### 4.3 Intervention-effect experiment

The intervention-effect experiment evaluated the environment-aware eHMI, projected eHMI, and auditory warning under the AT, rainy-weather, and non-yielding condition. The no-intervention condition was used as the reference category. Separate linear mixed-effects models with participant-specific random intercepts were estimated for perceived risk, objective risk, and pedestrian behavior. All models controlled for gender,age driving-experience category, and education category.

#### *4.3.1 Effects of intervention measures on risk perception*

The intervention effects on perceived risk are presented in Table 12. All three interventions significantly increased perceived risk in both crossing phases. In the first lane, auditory warning increased perceived risk by 0.722 points ($p < 0.001$), projected eHMI increased it by 0.889 points ($p < 0.001$), and environment-aware eHMI increased it by 0.333 points ($p < 0.01$). Projected eHMI produced the largest increase, followed by auditory warning and environment-aware eHMI.

A similar pattern was observed in the second lane. Auditory warning increased perceived risk by 0.463 points ($p < 0.001$), projected eHMI increased it by 0.870 points ($p < 0.001$), and environment-aware eHMI increased it by 0.296 points ($p < 0.05$). Thus, all three interventions enhanced pedestrians' awareness of the risk associated with the occluded far-side interaction, with projected eHMI again showing the strongest effect.

**Table 12** Intervention effects on perceived risk

| Risk perception | Intervention | Coefficient | SE | z-value | p-value |
|---|---|---|---|---|---|
| Lane 1 | Auditory warning | 0.722 | 0.119 | 6.07 | <0.001 |
| | Projected eHMI | 0.889 | 0.119 | 7.47 | <0.001 |
| | Environment-aware eHMI | 0.333 | 0.119 | 2.80 | 0.005 |
| Lane 2 | Auditory warning | 0.463 | 0.116 | 3.98 | <0.001 |
| | Projected eHMI | 0.870 | 0.116 | 7.49 | <0.001 |
| | Environment-aware eHMI | 0.296 | 0.116 | 2.55 | 0.011 |

*4.3.2 Effects of intervention measures on objective risk*

The significant intervention effects on the objective safety indicators are presented in Table 13. Projected eHMI was the only intervention associated with objective safety improvements.

**Table 13.** Effects of intervention measures on objective risk

| Objective risk | Intervention | Coefficient | SE | z-value | p-value |
|---|---|---|---|---|---|
| DST (Lane 1) | Projected eHMI | -0.142 | 0.052 | -2.75 | 0.006 |
| ITADV (Lane 2) | Projected eHMI | 0.255 | 0.117 | 2.18 | 0.029 |

In the first lane, projected eHMI reduced DST ($\beta = -0.142$, $p < 0.01$). Because lower DST values indicate a larger safety margin, this result suggests that projected eHMI improved objective safety during the interaction with the near-side AT.

In the second lane, projected eHMI increased ITADV ($\beta = 0.255$, $p < 0.05$). Since a higher ITADV indicates a larger temporal advantage relative to the far-side vehicle, projected eHMI also improved the objective safety margin during the occluded second-lane interaction.

Neither auditory warning nor environment-aware eHMI had a significant effect on DST or ITADV. Therefore, although all three interventions increased perceived risk, only projected eHMI was associated with significant improvements in both objective safety indicators.

*4.3.3 Effects of intervention measures on pedestrian behavior*

The significant behavioral effects are presented in Table 14. Environment-aware eHMI primarily influenced pedestrian behavior in the first lane. It increased first-lane waiting time ($\beta = 1.204$, $p < 0.001$), reduced first-lane crossing time ($\beta = -0.388$, $p < 0.05$), and increased first-lane vehicle observation time ($\beta = 0.613$, $p < 0.05$).

**Table 14.** Effects of intervention measures on pedestrian behavior

| Pedestrian behavior | Intervention | Coefficient | SE | z-value | p-value |
|---|---|---|---|---|---|
| Waiting time (Lane 1) | Environment-aware eHMI | 1.204 | 0.290 | 4.16 | <0.001 |
| Crossing time (Lane 1) | Environment-aware eHMI | -0.388 | 0.170 | -2.28 | 0.023 |
| Crossing time (Lane 2) | Projected eHMI | 0.222 | 0.078 | 2.83 | 0.005 |
| Vehicle observation time (Lane 1) | Projected eHMI | -0.952 | 0.282 | -3.37 | 0.001 |
| | Environment-aware eHMI | 0.613 | 0.282 | 2.17 | 0.030 |

These results indicate that pedestrians waited and observed the near-side vehicle for longer before crossing, but crossed the first lane more quickly after initiating movement. Environment-aware eHMI therefore encouraged a more cautious pre-crossing process, although the resulting behavioral changes were not accompanied by significant improvements in DST or ITADV.

Projected eHMI reduced first-lane vehicle observation time ($\beta = -0.952$, $p < 0.01$). This suggests that spatially localized information reduced the time needed to identify the relevant risk source. Projected eHMI also increased second-lane crossing time ($\beta = 0.222$, $p < 0.01$), indicating a slower and potentially more cautious traversal of the far-side lane.

Auditory warning did not significantly affect waiting time, crossing time, or vehicle observation time. Its influence was therefore concentrated on perceived risk rather than observable crossing behavior.

## 5. Discussion

### 5.1 How ATs shape pedestrian interaction risk

The results of this study suggest that ATs shape pedestrian interaction risk through a combination of vehicle appearance, motion-related constraints, and pedestrian behavioral responses. Their difference from passenger AVs is not limited to larger physical size. Rather, ATs introduce a distinct interaction context in which large body dimensions, stronger visual pressure, and reduced braking efficiency jointly change how pedestrians perceive risk and how objective safety margins are formed. This means that, in the tested scenario, pedestrian interaction with ATs may differ from passenger AVs interaction, suggesting a potentially distinct safety problem with its own risk structure (Nordhoff et al., 2025; Li et al., 2025).

The findings should be interpreted with ATs treated as an integrated interaction counterpart rather than as a simple combination of a truck platform and automated driving technology. Pedestrian interaction with AVs has been shown to depend jointly on vehicle automation, external appearance, driving behavior, communication capability, speed, and gap-related information, suggesting that pedestrians form crossing judgments from multiple vehicle and context cues rather than from a single isolated attribute (Dey et al., 2019; Velasco et al., 2019). In the present AT scenario, vehicle size, braking capability, automation status, communication ambiguity, and visibility effects were therefore considered as jointly perceived attributes of the approaching vehicle.

The perceived-risk results show that pedestrians were sensitive to both AT-related and weather-related risk cues. ATs significantly increased pedestrians' perceived risk, suggesting that pedestrians recognized the potential danger associated with large vehicle size, stronger visual pressure, and more severe possible conflict consequences. This finding is consistent with previous heavy-vehicle safety studies showing that larger vehicles tend to impose greater visual pressure and are associated with higher injury severity in pedestrian conflicts (Tyndall, 2024; Edwards and Leonard, 2022). Rainy conditions also increased perceived risk, indicating that pedestrians treated adverse weather as a meaningful risk cue rather than merely as a visual background. In this sense, pedestrians' subjective risk judgment appeared to be strongly driven by salient and directly observable cues, including vehicle type and environmental condition.

The objective-risk results, however, present a more nuanced pattern. ATs were associated with a larger temporal safety margin under some conditions, which should not be interpreted as evidence that ATs are inherently safer. A more plausible explanation is risk compensation: when pedestrians perceived ATs as more threatening, they tended to adopt more conservative crossing timing, thereby preserving greater safety redundancy. This interpretation is consistent with the view that road users may adjust their behavior when they perceive a situation as more dangerous (Graham et al., 1983). Nevertheless, this compensation was not stable. The interaction between vehicle type and weather condition significantly weakened the temporal safety margin associated with ATs, suggesting that rainy conditions reduced the effectiveness of pedestrians' cautious behavior by further constraining braking efficiency. Wet pavement can make deceleration and stopping more difficult, and this effect is especially relevant for trucks because of their greater mass and stronger inertia (Wang et al., 2023; Jiang et al., 2024; Jung et al., 2014). Therefore, weather did not merely increase perceived danger; it also altered the objective safety margin by changing vehicle motion constraints in AT interactions.

An important implication is that perceived risk and objective risk were not governed by exactly the same structure. Vehicle type and weather condition both directly increased perceived risk, but the combined effect of ATs and rainy conditions was not reflected in the perceived-risk model. By contrast, this combined effect was significant for the objective-risk outcome. This suggests that pedestrians could recognize the apparent danger of ATs and rainy weather separately, but may not fully internalize how their combination changes the actual safety margin through reduced braking efficiency. In other words, pedestrians may respond clearly to visible and intuitive risk cues, such as vehicle size and weather, while the dynamic consequences of reduced braking efficiency are less directly perceptible. This mismatch helps explain why higher perceived risk can coexist with a more favorable objective safety margin in some conditions, and why that margin can still be reduced when vehicle dynamics become more constrained.

### 5.2 How ATs affect pedestrian interaction with other vehicles through occlusion

The findings suggest that the risk associated with ATs is not limited to their role as direct interaction partners for pedestrians. Because of their large body size, ATs can also reshape the surrounding interaction environment by obstructing pedestrians' view of other traffic participants. In this sense, an AT may influence not only how pedestrians interact with the truck itself, but also how they judge and respond to vehicles beyond the truck. This indirect influence through occlusion represents an important distinction between ATs and passenger AVs.

The results for the second lane support this interpretation. Under the AT condition, pedestrians' ITADV was significantly reduced, indicating that the time advantage retained when interacting with the far-side vehicle was compressed and that objective interaction risk increased. The mediation analysis further showed that second-lane entry time played a partial mediating role, suggesting that ATs affected subsequent risk partly by changing pedestrians' temporal position before they encountered the far-side vehicle. However, after controlling for second-lane entry time, the direct effect of vehicle type on ITADV remained significant. This indicates that the increase in second-lane risk cannot be explained solely by a timing-related pathway. Rather, ATs also exerted a remaining direct influence on subsequent interaction safety.

This remaining effect is most plausibly related to occlusion. Because ATs have larger bodies and occupy a wider visual corridor, pedestrians' ability to identify the position, speed, and arrival timing of the far-side vehicle may be weakened after the direct interaction with the near-side vehicle (Zhu et al., 2021). Under such conditions, pedestrians are not only deciding whether sufficient time remains for crossing; they are also making this judgment under incomplete information. Therefore, the elevated subsequent risk reflects both restricted information acquisition and compressed time advantage. The mediation analysis is important because it separates the timing-related pathway from the remaining direct effect, thereby showing that AT-related risk is not transmitted only through changes in pedestrian movement timing, but also through the obstruction of far-side risk recognition.

This finding is consistent with research on occlusion risk and restricted visibility in heavy-vehicle contexts. The large body size of trucks can enlarge occluded areas and increase the difficulty vulnerable road users face when judging surrounding traffic conditions (Jansen and Varotto, 2022; Schindler and Piccinini, 2021). The present study extends this understanding to AT scenarios by showing that the influence of large vehicle bodies is not confined to the direct conflict between the truck and the pedestrian. It may also spill over into pedestrians' subsequent interaction with other vehicles. Compared with passenger AVs studies, which often focus on how vehicles communicate their own intentions to pedestrians, the present findings suggest that AT scenarios require additional attention to how the vehicle affects pedestrians' recognition of other risk sources.

### 5.3 Intervention effectiveness

The results indicate that, in AT scenarios, the key role of intervention measures is not only to make pedestrians aware that risk exists, but also to help them identify where the risk is located. Under occlusion conditions, a warning that merely increases general alertness may be insufficient if it does not support pedestrians in judging the potential conflict area or locating the risk source. From this perspective, the different effects of projected eHMI, environment-aware eHMI, and auditory warning can be understood in terms of their information carrier, spatial directivity, and pathway of influence.

The advantage of projected eHMI lies in its ability to bind warning information directly to the potential conflict area. The results showed that projected eHMI significantly reduced DST and increased ITADV, while being accompanied by more favorable behavioral adjustment. This suggests that projected information did not simply inform pedestrians that risk existed, but helped them localize the risk and adjust their crossing process by placing the warning near the relevant conflict space. This feature is particularly important in AT scenarios, where occlusion may prevent pedestrians from accurately identifying the position of the risk source and its spatiotemporal relationship to themselves, even when they are already aware that the situation is risky (Tabone et al., 2023). In this sense, projected eHMI appears to address the risk-localization problem more directly, making it potentially better aligned with the safety demands of AT occlusion scenarios.

By contrast, although environment-aware eHMI significantly increased perceived risk and altered pedestrians' waiting and observation behavior, it did not produce a significant improvement in either first-lane DST or second-lane ITADV. A possible explanation is that environment-aware eHMI presents information on the surface of the near-side vehicle, which may direct pedestrians' attention toward the current interaction object rather than the broader conflict space. Although this display can emphasize the near-side AT and its warning message, it does not necessarily indicate the location of the subsequent risk source. In AT occlusion scenarios, where the critical hazard may arise from a visually blocked far-side vehicle (Chen et al., 2025), vehicle-mounted information may further concentrate pedestrians' attention on the truck itself rather than guide them toward the potential conflict area. This may explain why environment-aware eHMI changed pre-crossing behavior but did not consistently translate into measurable objective safety benefits.

Auditory warning showed a different pattern. The results showed that auditory warning significantly increased pedestrians' perceived risk, indicating that it was effective in risk arousal. However, its effects on objective safety indicators were not stable, suggesting that increasing alertness alone may not be sufficient to address the central problem in AT scenarios. The limitation of auditory warning lies in its weak spatial directivity. It can inform pedestrians that a risk exists, but it provides limited support for determining where that risk is located. Under occlusion conditions, pedestrians may need not only a warning signal, but also assistance in locating and confirming the obscured risk source (Ulrich et al., 2014). Therefore, auditory warning may be more suitable as a supplementary risk-arousal method, while its effectiveness is more limited when spatial risk recognition is required.

### 5.4 Practical implications

The findings of this study suggest that the safety challenges associated with ATs require more targeted design and deployment strategies than those of passenger AVs. Their interaction risk arises not only from the vehicle's own characteristics, but also from their influence on pedestrians' risk recognition and subsequent traffic interactions.

First, greater caution may be warranted when deploying ATs in high-risk scenarios, particularly under conditions such as rain that affect braking capability. The results indicate that reduced braking efficiency under rainy conditions further weakens the safety margin in pedestrian-ATs interactions. Accordingly, such scenarios may deserve priority in practical risk identification and management. More specifically, under rainy weather, low-adhesion pavement, or reduced-visibility conditions, ATs may benefit from more conservative operating strategies, including lower approach speeds, earlier deceleration, and larger yielding and stopping margins (Hsu and Jones, 2017; Zhang et al., 2023). In unsignalized crossing areas, especially in multi-lane settings, geofencing or scene-recognition mechanisms could also be used to automatically switch vehicles into a high-caution mode, thereby limiting overly aggressive approaches toward pedestrians. At the operational level, these scenarios may also be treated as priority monitoring areas where additional warning devices or higher-level interaction support systems are considered (Fwa, 2017; Jiang et al., 2024).

Second, from the perspective of traffic safety management, the indirect effects of ATs on interactions between pedestrians and other vehicles should also be taken into account, particularly the transmission of risk caused by occlusion. Traditional pedestrian-vehicle safety assessment has focused primarily on the direct relationship between a single vehicle and a single pedestrian. However, the present study suggests that ATs may also alter pedestrians' recognition and judgment of other approaching vehicles through occlusion, thereby affecting subsequent interaction safety. This implies that the risk associated with ATs is not merely local, but may reshape the interaction process more broadly. Therefore, in road safety management and in the testing and evaluation of automated vehicles, it may not be sufficient to assess only whether the vehicle itself decelerates or yields. It may also be important to consider whether the vehicle obstructs pedestrians' recognition of subsequent risk sources during the interaction. In scenarios where this type of risk is more pronounced, management priorities may include reducing information loss under occlusion, for example by improving the vehicle's ability to provide risk-related information in complex interactions, strengthening warning support for subsequent risk sources, and incorporating into safety evaluation whether occlusion changes pedestrians' subsequent decisions and safety margins. Such considerations may help provide a more complete response to the safety issues raised by AT scenarios (Tomasch and Smit, 2023; Schindler and Piccinini, 2021).

Finally, with respect to interface design, priority may be given to helping pedestrians identify risk sources and potential conflict areas. The results of this study indicate that, in ATs occlusion scenarios, the key difficulty faced by pedestrians is not simply a lack of awareness of vehicle presence, but rather the difficulty of determining where the actual risk is located under information-constrained conditions. Therefore, compared with interfaces that merely display vehicle status or intention on the vehicle surface, designs with stronger spatial directivity may be more likely to improve objective safety. For ATs, interaction interfaces may need to address the question of where the potential risk is located. Especially when the near-side vehicle has already created occlusion, more effective warning strategies may be those that establish a direct connection between the information and the potential conflict area, so that pedestrians can form a clearer judgment of risk location before entering the subsequent lane, rather than continuing to focus their attention on the near-side vehicle itself. In light of the present findings, the advantage of projected eHMI may lie in its ability to transfer risk information from the vehicle surface to the conflict space, potentially aligning more closely with the interaction demands of AT scenarios (Chen et al., 2025; Tabone et al., 2023).

## 6. Conclusions

This study investigated pedestrian-AT interaction risk and occlusion-targeted interventions in an unsignalized multi-lane crossing scenario using a controlled VR experiment. A two-stage design was adopted: the first stage examined risk formation under different combinations of weather condition, near-side vehicle type, and far-side vehicle interaction strategy, and the second stage evaluated three interventions under the representative high-risk scenario: environment-aware eHMI, projected eHMI, and auditory warning.

The results show that ATs affected pedestrian interaction risk through both direct pedestrian-truck interaction and occlusion-induced risk. In the first lane, ATs increased perceived risk but also induced more conservative crossing behavior, producing a larger temporal safety margin under specific conditions. However, this compensation was weakened under rainy conditions, suggesting that reduced braking efficiency can offset the safety redundancy generated by pedestrian caution. In the second lane, ATs reduced pedestrians' ITADV, and mediation analysis indicated that this effect was related not only to second-lane entry timing but also to the occlusion caused by the large truck body.

The intervention results show that all three measures increased pedestrians' risk awareness, but their effects on behavior and objective safety differed. Projected eHMI showed the most favorable overall performance by improving DST, significantly increasing ITADV, and supporting behavioral adjustment consistent with risk localization. Environment-aware eHMI increased perceived risk and changed pedestrian behavior but did not produce consistently favorable objective safety outcomes, while auditory warning mainly enhanced risk awareness with limited evidence of behavioral or objective safety improvement. These findings suggest that pedestrian-oriented interventions for ATs should not only alert pedestrians to potential danger but also help them locate the risk source.

Several limitations should be acknowledged. First, although VR enabled safe and controlled manipulation of AT-related risk factors, it cannot fully reproduce the complexity of real-world traffic. Second, the participants were mainly young university students, which may limit the generalizability of the findings to older pedestrians or other road-user groups. Third, this study focused on a specific unsignalized two-lane crossing scenario. Future research should examine more diverse road environments, participant groups, vehicle configurations, speeds, and combined intervention strategies to further validate occlusion-targeted safety support in pedestrian-AT interaction.

**Authorship contribution statement**

**Yun Ye:** Conceptualization, Data curation, Formal analysis, Project administration, Investigation, Methodology, Validation, Funding acquisition, Writing-original draft. **Yuan Che:** Methodology, Data curation, Formal analysis, Investigation, Visualization, Writing-original draft. **S.C. Wong:** Conceptualization, Writing-review & editing, Funding acquisition. **Stergios-Aristoteles Mitoulis:** Supervision, Writing-review & editing. **Haoyang Liang:** Conceptualization, Validation, Resources, Funding acquisition, Writing-review & editing.

**Declaration of the use of AI**

During the preparation of this work the authors used ChatGPT in order to improve the readability and language of the manuscript. After using this tool/service, the authors reviewed and edited the content as needed and take full responsibility for the content of the publication.

**Ethical statement**

Prior to the commencement of the study, ethical approval for the study protocol was granted by Science and Technology Ethics Committee of Tongji University with the approval number: tjdxsr2024041.

**Data availability statement**

Data will be made available on reasonable request. The accessible URL is https://osf.io/tzyqs/overview.

**Declaration of competing interest**

The authors declare that they have no known competing financial interests or personal relationships that could have appeared to influence the work reported in this paper.

# Appendix

**Appendix A1.** Virtual reality sickness questionnaire (VRSQ)

| VRSQ symptom | Oculomotor | Disorientation |
|---|---|---|
| 1. General discomfort | O | |
| 2. Fatigue | O | |
| 3. Eyestrain | O | |
| 4. Difficulty focusing | O | |
| 5. Headache | | O |
| 6. Fullness of head | | O |
| 7. Blurred vision | | O |
| 8. Dizzy (eyes closed) | | O |
| 9. Vertigo | | O |

**Appendix A2.** Condition-wise descriptive statistics for Stage 1 risk formation component

| Weather | Near-side vehicle type | Far-side strategy | DST | ITADV | Risk perception (Lane 1) | Risk perception (Lane 2) |
|---|---|---|---|---|---|---|
| Rainy | AV | Non-yielding | -0.20 ± 0.46 | 1.58 ± 2.42 | 5.50 ± 0.77 | 7.52 ± 0.57 |
| Rainy | AV | Yielding | -0.17 ± 0.46 | 2.34 ± 1.67 | 5.60 ± 0.72 | 6.93 ± 0.70 |
| Rainy | AT | Non-yielding | -0.20 ± 1.88 | 0.57 ± 1.76 | 8.57 ± 0.60 | 8.54 ± 0.79 |
| Rainy | AT | Yielding | 0.01 ± 0.24 | 0.18 ± 0.66 | 8.44 ± 0.72 | 7.70 ± 0.57 |
| Sunny | AV | Non-yielding | -0.06 ± 0.17 | 1.51 ± 2.48 | 4.54 ± 0.84 | 7.22 ± 0.95 |
| Sunny | AV | Yielding | -0.09 ± 0.35 | 2.31 ± 2.12 | 4.46 ± 0.69 | 6.48 ± 0.64 |
| Sunny | AT | Non-yielding | -0.43 ± 1.73 | 0.61 ± 1.64 | 7.50 ± 0.69 | 8.33 ± 0.61 |
| Sunny | AT | Yielding | -0.29 ± 1.65 | 0.42 ± 0.82 | 7.59 ± 0.69 | 7.39 ± 0.63 |

**Appendix A3.** Condition-wise descriptive statistics for Stage 2 intervention-effect component

| **Intervention** | **Risk perception (Lane 1)** | **Waiting time (Lane 1)** | **Crossing time (Lane 1)** | **Vehicle observation time (Lane 1)** | **DST** | **Risk perception (Lane 2)** | **Waiting time (Lane 2)** | **Crossing time (Lane 2)** | **Vehicle observation time (Lane 2)** | **ITADV** |
|---|---|---|---|---|---|---|---|---|---|---|
| None | 8.44 ± 0.72 | 8.34 ± 2.00 | 2.97 ± 1.32 | 4.98 ± 2.06 | 0.01 ± 0.24 | 7.70 ± 0.57 | 2.66 ± 1.25 | 2.46 ± 0.18 | 0.64 ± 0.74 | 0.18 ± 0.66 |
| Auditory warning | 9.17 ± 0.69 | 8.35 ± 2.10 | 2.71 ± 1.07 | 4.81 ± 2.11 | 0.00 ± 0.28 | 8.17 ± 0.75 | 2.89 ± 1.23 | 2.48 ± 0.25 | 0.66 ± 0.78 | 0.21 ± 0.72 |
| Projected eHMI | 9.33 ± 0.64 | 8.41 ± 2.84 | 2.93 ± 1.10 | 4.03 ± 2.38 | -0.13 ± 0.60 | 8.57 ± 0.57 | 2.56 ± 1.21 | 2.68 ± 0.77 | 0.68 ± 0.91 | 0.44 ± 1.03 |
| Environment-aware eHMI | 8.78 ± 0.69 | 9.54 ± 2.48 | 2.58 ± 0.89 | 5.59 ± 2.64 | 0.07 ± 0.12 | 8.00 ± 0.85 | 2.68 ± 1.21 | 2.45 ± 0.23 | 0.57 ± 0.81 | 0.17 ± 0.49 |

**Appendix A4.** Total-effect model (Dependent variable: ITADV, path c)

| **Variable** | **Coefficient** | **SE** | **z-value** | **p-value** | **95% CI** |
|---|---|---|---|---|---|
| Autonomous truck | -1.494 | 0.157 | -9.51 | $<$0.001 | [-1.802, -1.186] |
| Rainy weather | -0.046 | 0.157 | -0.29 | 0.768 | [-0.354, 0.262] |
| Non-yielding far-side strategy | -0.246 | 0.157 | -1.57 | 0.118 | [-0.554, 0.062] |
| Gender | 0.502 | 0.316 | 1.59 | 0.112 | [-0.117, 1.120] |
| Age | -0.191 | 0.104 | -1.83 | 0.067 | [-0.395, 0.014] |
| Driving experience | | | | | |
| 0 years | | | Reference | | |
| Less than 1 year | -0.379 | 0.42 | -0.90 | 0.366 | [-1.202, 0.443] |
| 1–3 years | -0.016 | 0.422 | -0.04 | 0.969 | [-0.844, 0.811] |
| 3–More than 5 years | -0.166 | 0.429 | -0.39 | 0.699 | [-1.006, 0.674] |
| >More than 5 years | 0.406 | 0.544 | 0.75 | 0.456 | [-0.661, 1.473] |
| Education level | | | | | |
| Bachelor's degree or lower | | | Reference | | |
| Master's degree or higher | 0.06 | 0.367 | 0.16 | 0.869 | [-0.658, 0.779] |
| Intercept | 6.465 | 2.526 | 2.56 | 0.010 | [1.514, 11.416] |

**Appendix A5.** Mediator model (Dependent variable: $T_{entry}$ , path a)

| Variable | Coefficient | SE | z-value | p-value | 95% CI |
|---|---|---|---|---|---|
| Autonomous truck | 1.004 | 0.134 | 7.51 | <0.001 | [0.742, 1.265] |
| Rainy weather | 0.272 | 0.134 | 2.04 | 0.042 | [0.010, 0.534] |
| Non-yielding far-side strategy | -0.068 | 0.134 | -0.51 | 0.612 | [-0.330, 0.194] |
| Gender | -1.262 | 0.552 | -2.29 | 0.022 | [-2.343, -0.181] |
| Age | 0.245 | 0.182 | 1.35 | 0.178 | [-0.112, 0.603] |
| Driving experience | | | | | |
| 0 years | | | Reference | | |
| Less than 1 year | -0.328 | 0.734 | -0.45 | 0.655 | [-1.766, 1.110] |
| 1–3 years | 0.021 | 0.738 | 0.03 | 0.978 | [-1.426, 1.467] |
| 3–More than 5 years | -0.307 | 0.749 | -0.41 | 0.682 | [-1.775, 1.162] |
| >More than 5 years | -0.229 | 0.952 | -0.24 | 0.81 | [-2.095, 1.636] |
| Education level | | | | | |
| Bachelor's degree or lower | | | Reference | | |
| Master's degree or higher | -0.010 | 0.641 | -0.01 | 0.988 | [-1.266, 1.247] |
| Intercept | 0.346 | 4.411 | 0.08 | 0.938 | [-8.300, 8.991] |

**Appendix A6.** Direct-effect model (Dependent variable: ITADV, paths b and c')

| Variable | Coefficient | SE | z-value | p-value | 95% CI |
|---|---|---|---|---|---|
| Autonomous truck | -1.089 | 0.152 | -7.17 | <0.001 | [-1.387, -0.792] |
| Second-lane entry time | -0.403 | 0.041 | -9.87 | <0.001 | [-0.484, -0.323] |
| Rainy weather | 0.064 | 0.147 | 0.43 | 0.665 | [-0.224, 0.351] |
| Non-yielding far-side strategy | -0.273 | 0.146 | -1.87 | 0.062 | [-0.560, 0.013] |
| Gender | -0.007 | 0.247 | -0.03 | 0.976 | [-0.491, 0.476] |
| Age | -0.092 | 0.08 | -1.14 | 0.254 | [-0.249, 0.066] |
| Driving experience | | | | | |
| 0 years | | | Reference | | |
| Less than 1 year | -0.512 | 0.321 | -1.60 | 0.111 | [-1.141, 0.117] |
| 1–3 years | -0.008 | 0.322 | -0.02 | 0.981 | [-0.640, 0.624] |
| 3–More than 5 years | -0.290 | 0.328 | -0.88 | 0.377 | [-0.932, 0.352] |
| >More than 5 years | 0.313 | 0.416 | 0.75 | 0.451 | [-0.502, 1.129] |
| Education level | | | | | |
| Bachelor's degree or lower | | | Reference | | |
| Master's degree or higher | 0.056 | 0.28 | 0.2 | 0.84 | [-0.493, 0.605] |
| Intercept | 6.604 | 1.931 | 3.42 | 0.001 | [2.820, 10.389] |

**Appendix A7.** Random effects and model fit

| Model | Dependent variable | Random-intercept variance | Residual variance | Log likelihood | Wald ( $\chi^2$ ) | LR test versus linear model |
|---|---|---|---|---|---|---|
| Total-effect | ITADV | 0. 475 | 2. 667 | −848. 761 | 102.23, ( $p < 0.001$ ) | 22.97, ( $p < 0.001$ ) |
| Mediator | $T_{entry}$ | 2. 229 | 1. 929 | − 817. 716 | 70,27, ( $p < 0.001$) | 206.13, ( $p < 0.001$) |
| Direct-effect | ITADV | 0. 183 | 2. 307 | − 806. 828 | 220.85, ( $p < 0.001$) | 6.41, ( $p < 0.01$) |